\documentclass[namedreferences]{SolarPhysics}
\usepackage{epsfig}

\newcommand{\be}{\begin{equation}}
\newcommand{\ee}{\end{equation}}
\newcommand{\beq}{\begin{eqnarray}}
\newcommand{\eeq}{\end{eqnarray}}

\usepackage{graphicx} 
\DeclareMathAlphabet{\mathbfl}{OML}{cmm}{b}{it} 

\newcommand{\DS}{\ensuremath{\displaystyle}}

\begin{document}
\bibliographystyle{aa}

\begin{article}
\begin{opening}
\title{Magnetic Stereoscopy}
\author{T. \surname{Wiegelmann}\email{wiegelmann@mps.mpg.de},
B. \surname{Inhester}}
\institute{Max-Planck-Institut f\"ur Sonnensystemforschung,
Max-Planck-Strasse 2, 37191 Katlenburg-Lindau, Germany}


\runningtitle{Stereoscopy}
\runningauthor{Wiegelmann et al.}
\date{Solar Physics, Vol. 236, 25-40, 2006}
\begin{abstract}
 The space mission STEREO will provide images from two
 viewpoints. An important aim of the STEREO mission is
 to get a 3D view of the solar corona.
 We develop a program for the stereoscopic reconstruction of
 3D coronal loops from images taken with the two STEREO spacecraft.
 A pure geometric triangulation of coronal features leads to ambiguities
 because the dilute plasma emissions complicates the association of features
 in image 1 with features in image 2.  As a consequence of these problems
 the stereoscopic reconstruction is not unique and multiple solutions occur.
 We demonstrate how these ambiguities can be resolved with the help
 of different coronal magnetic field models (potential, linear and non-linear
 force-free fields). The idea is that, due to the high conductivity in
 the coronal plasma, the emitting plasma outlines the magnetic field lines.
 Consequently the 3D coronal magnetic field provides a proxy for the
 stereoscopy which allows to eliminate inconsistent configurations.
 The combination of stereoscopy and magnetic modelling is more powerful
 than one of these tools alone. We test our method with the help of
 a model active region and plan to apply it to the solar case as soon as
 STEREO data become available.
\end{abstract}
\keywords{STEREO, magnetic fields, extrapolations}
\end{opening}

\section{Introduction}
\label{sec1} The forthcoming space mission STEREO
 (Solar TErrestrial RElations Observatory) will observe the Sun
simultaneously from two viewpoints. One aim of the mission is to reconstruct the
solar corona in three dimensions (see, e.g., \opencite{schmidt:etal96};
\opencite{gary:etal98} for an overview). To do so we have to develop tools for the
stereoscopic reconstruction of the 3D corona from  two STEREO-images. A
triangulation method using the solar rotation has been applied to Skylab images by
\inlinecite{berton:etal85} and \inlinecite{batchelor94}.
\citeauthor{aschwanden:etal99}(\citeyear{aschwanden:etal99},
\citeyear{aschwanden:etal00}) and \inlinecite{portier-fozzani:etal01} used SOHO data
and the rotation of the Sun for a stereoscopic reconstruction. Using the rotation of
the Sun and taking images a few days apart allows of course only the reconstruction
of features with remain stationary within this time. \inlinecite{aschwanden:etal99}
made a fit of the observed loop structures to sections of circles and allowed for a
time dependence of some of the fit parameters. The method was called {\it dynamic
stereoscopy} and used the assumption of circular coronal loops as a priori
information. \inlinecite{wiegelmann:etal02} demonstrated how the 3D loops published
in \cite{aschwanden:etal99} and photospheric magnetic field measurements can be used
to compute a corresponding coronal magnetic field model within the linear force-free
model. A basic assumption was that due to the high conductivity the emitting coronal
plasma also outlines the magnetic field. The alignment of coronal plasma and
magnetic field lines has also be used directly with 2D images from (Yokoh/SXT) in
\cite{carcedo:etal03} and from (SOHO/EIT) in \cite{marsch:etal04} to compute the
optimal coronal magnetic field within the linear force-free assumption.
\inlinecite{wiegelmann:etal05b} used linear and non-linear force-free magnetic field
models for the identification of coronal loops in EUV images.

The aim of this work is to develop a tool for the stereoscopic reconstruction of
coronal features (mainly closed loops in active regions) from two viewpoints. We
test the quality of our reconstruction tools with the help of a model active region.
Pure geometric stereoscopy leads to multiple solutions, mainly because the faint
coronal plasma does not allow a clear association of features in both Stereo-images
with each other. Classical stereoscopy works best for objects with clear edges in
images with high contrast. Unfortunately this is not the case in the solar corona
where the plasma structures are very faint and diffuse, e.g., visible loops in
high-resolution TRACE images are often a superposition of several individual loops
(\citeauthor{schrijver:etal99}, \citeyear{schrijver:etal99},
\citeyear{schrijver:etal04}).

We demonstrate how a suitable coronal magnetic field model can be used to find the
association of loops in the STEREO images and thereby remove  ambiguities in the
stereoscopic reconstruction. The method also tell us, how well  the assumed coronal
magnetic field model aligns observed loops in both images and computes (for linear
force-free models) the optimal value of $\alpha$. The tools are planned for use
within the STEREO-mission.

\section{Stereoscopy tools}
\subsection{Geometric Stereoscopy}
\label{Geometric_Stereoscopy} By geometric stereoscopy we understand  a 3D
reconstruction from two images, e.g., from the projections as shown in Figure
\ref{fig2}. As real STEREO data are not yet available, we test our method with the
help of a model active region, as described in appendix \ref{appendix1}. Using a
model active region helps us (because we know the true solution) to check how
accurate our stereoscopic reconstruction tools work. Here we try to reconstruct the
four loops in 3D from the artificial images shown in Figure \ref{fig2}. For real
observed images from e.g., the two STEREO-spacecraft one needs to get the loops (or
after \opencite{aschwanden05} curvi-linear 1D features.) first from the 2D EUV
images by feature tracking method. Several methods have been proposed for this aim,
e.g., the brightness-gradient method and the oriented-connectivity method
 (See \opencite{aschwanden05}; \opencite{lee:etal06} for an overview). Here we concentrate on
 the 3D reconstruction and assume that the two EUV images have been
 transformed into curvi-linear 1D features. In the following we call these
 observed elongated structures in EUV-images simply {\it loops}.
In contrast field lines are 3D curves derived from magnetic field models.
The assumption is that loops and projections of field lines are aligned.

We make a back projection of the 4 images into the original 3D box. Geometric
stereoscopy works well for solid objects with well distinguishable edges. If one has
correctly identified two related points in both images, a computation of the 3D
location of the point is straight forward. One just has to calculate the point of
intersection along the line-of-sight of both images. Unfortunately the situation is
more complicated for the solar corona. Coronal loops are faint elongated objects and
often have no clearly visible edges. It is not clear a priori which points along a
loop projection in image 1 belong to which points along the same loop from another
viewpoint in image 2. The situation becomes even worse for multiple loops which are
close together in the images. Here it is not always possible to distinguish which
loops from image 1 correspond to which loops in image 2. For a stereoscopic
reconstruction in such a situation we compute the intersection points of all
identified loop points in image 1 with all identified loop points in the second
image \footnote{It is sufficient to search for intersection points which are on the
same epipolar line, because points on different epipolar lines do not intersect.}.
The 3D reconstruction is not unique, however, because a pixel in one image usually
intersects with more than one pixel in the other image. An example of a pure
geometric stereoscopic reconstruction is shown in Figure \ref{fig1} b). The {\it
black} pixels mark the 3D intersection of the 3D reconstruction, the {\it yellow}
dotted lines mark the original loops. One can see that geometric stereoscopy finds
the correct 3D locations of the loops and reconstructs the original loop structure,
but also several ghost features occur, which are not related to any real loop. The
challenge is to identify which intersections are real and correspond to magnetic
loops on the Sun and to eliminate the ghost points.
\subsection{Magnetic modelling}
\label{magnetic_modelling} While the blurring and the line-of-sight character of the
coronal images complicate the interpretation, the coronal plasma has the nice
feature that the plasma emission outlines the magnetic field. This is a consequence
of the high conductivity of the coronal plasma. Outlining means that the loops
visible in EUV images also represent projections of the magnetic field lines.
Consequently the 3D reconstruction of coronal plasma loops and 3D magnetic field
lines are associated with each other. Unfortunately the 3D coronal magnetic field
cannot be measured directly and one has to extrapolate the field from photospheric
measurements into the corona. The extrapolation depends on assumptions regarding the
coronal plasma, in particular the electric current density. In the low and middle
corona the magnetic pressure dominates over the plasma pressure ($\beta \ll 1$ )
which allows to use force-free magnetic field models. Force-free magnetic fields
have to obey the equations
\begin{eqnarray}
(\nabla \times {\bf B }) \times{\bf B} & = & {\bf 0},
\label{forcefree} \\
\nabla \cdot{\bf B}    & = &         0.
\label{solenoidal-ff}
\end{eqnarray}
which are equivalent to
\begin{eqnarray}
(\nabla \times {\bf B }) & = & \alpha {\bf B}, \\
{\bf B } \cdot \nabla \alpha & = & 0.
\end{eqnarray}

The in general non-linear force-free field model
\cite{sakurai81,amari:etal99,wheatland:etal00,yan:etal00,regnier:etal02,wiegelmann:etal03,wiegelmann04,wheatland04,valori:etal05,wiegelmann:etal06,inhester:etal06}
has potential fields (no currents, e.g., \opencite{semel67}) and linear force-free
fields (current proportional to the magnetic field with a global constant $\alpha$,
e.g., \opencite{chiu:etal77};\opencite{seehafer78}) as subclasses. Potential and
linear force-free fields only need the line-of-sight magnetic field as input, which
is observed routinely from e.g., SOHO/MDI and NSO/Kitt Peak. Non-linear force-free
fields are mathematically more challenging to compute and require photospheric
vector magnetograms as input. Such data contain high noise and ambiguities in the
transverse magnetic field component and currently operating vectormagnetographs
(e.g., NAO/SFT, VTT in Tenerife and IVM in Hawaii) have a limited field of view. The
observational shortage of vectormagnetograph data will improve however in the near
future with the forthcoming missions Solar-B, SOLIS and SDO. The non-linear
force-free approach describes the magnetic field in active regions more accurately
than potential and linear force-free fields (see \opencite{wiegelmann:etal05}).
\subsubsection{Potential and linear force-free fields.}
We use the \cite{seehafer78} method to calculate potential and  linear force-free
fields.  The method requires  a photospheric line-of-sight magnetogram (e.g., from
SOHO/MDI) as input. The Seehafer solution is computed on a rectangular grid $0 \dots
L_x$ and $0 \dots L_y$ and contains the free force-free parameter $\alpha$, which
cannot be evaluated from the observed line-of-sight magnetic field. To normalize
$\alpha$ we choose the harmonic mean $L$ of $L_x$ and $L_y$ defined by
$\frac{1}{L^2}=\frac{1}{2}\left(\frac{1}{L_x^2}+\frac{1}{L_y^2}\right)$.  The
force-free parameter is limited by $-\sqrt{2} \pi < \alpha L < \sqrt{2} \pi.$
Potential fields correspond to $\alpha=0$. (See \opencite{seehafer78} for details.)
\subsubsection{Non-linear force-free fields.}
We solve  Eqs. (\ref{forcefree}) and (\ref{solenoidal-ff}) with the help of an
optimization principle  as proposed by \inlinecite{wheatland:etal00} and generalized
by \inlinecite{wiegelmann:etal03a}; \inlinecite{wiegelmann04}:
\begin{equation}
L=\int_{V}  w(x,y,z) \, \left[B^{-2} \, |(\nabla \times {\bf B}) \times {\bf
B}|^2 +|\nabla \cdot {\bf B}|^2\right] \, d^3x
\label{defL1},
\end{equation}
where $w(x,y,z)$ is a weighting function.  It is obvious that (for $w>0$) the
force-free Eqs. (\ref{forcefree}-\ref{solenoidal-ff}) are fulfilled when
L equals zero.    As an initial
configuration we compute a potential magnetic field in the computational box.
 As the next step we use photospheric
vector magnetic field data to prescribe the bottom boundary (photosphere) of the
computational box. On the lateral and top boundaries the field is chosen from the
potential field above.  We iterate for the magnetic field inside the computational
box by minimizing Eq. (\ref{defL1}). The weighting function $w$ equals 1 everywhere
in the computational box except in a boundary layer of $16$ points towards the
lateral and top boundary of the computational box, where $w$ decreases smoothly to 0
with a cosine function. (See \opencite{wiegelmann04} for details of our
implementation of the non-linear force-free optimization principle.)

\begin{figure}
\begin{picture}(16,20)
  \put( 0   ,18){\framebox(3.5,2){image 1}}
  \put( 1.75,18){\vector(0,-1){1}}
  \put( 4   ,18){\framebox(3.5,2){image 2}}
  \put( 5.75,18){\vector(0,-1){1}}
  \put( 8,18){\framebox(3.5,2){\parbox{3.5cm}{\begin{center}
      B-surface\\[-2mm]data\end{center}}}}
  \put(10,   18){\vector(0,-1){1.5}}
  \put(12   ,18){\framebox(3.5,2){\parbox{3.5cm}{\begin{center}
      poss. add.\\[-2mm]parameters,\\[-2mm]
      e.g., $\alpha$\end{center}}}}
  \put(13.5 ,18){\vector(0,-1){1.5}}

  \put( 0,   14){\framebox(3.5,3){\parbox{3.5cm}{\begin{center}
      segment\\[-2mm] image 1\\[-2mm]into loops\\
      $\ell_1=1\dots n_1$\end{center}}}}
  \put( 1.75,14){\vector(0,-1){2.5}}
  \put( 4,14){\framebox(3.5,3){\parbox{3.5cm}{\begin{center}
      segment\\[-2mm] image 2\\[-2mm]into loops\\
      $\ell_2=1\dots n_2$\end{center}}}}
  \put( 5.75,14){\vector(0,-1){2.5}}
  \put(9.25,14.5){\framebox(5,2){\parbox{5cm}{\begin{center}
      calculate coronal\\ B-field model\end{center}}}}
  \put(11.75,14.5){\vector(0,-1){4}}

  \put( 1.75, 13){\makebox(2.5,0){forall $\ell_1$}}
  \put( 5.75, 13){\makebox(2.5,0){forall $\ell_2$}}

  \put(-0.5,4){\dashbox(17,8){}}

  \put( 0,   7.5){\framebox(3.5,4){\parbox{3.5cm}{\begin{center}
      calculate\\ $C_{\ell_1}(b)=$\\[1mm]
      $\frac{\DS\mathrm{area}(\ell_1,b)}
            {\DS\mathrm{length}(\ell_1)}$\end{center}}}}
  \put( 1.75,7.5){\line(0,-1){0.5}}
  \put( 4,   7.5){\framebox(3.5,4){\parbox{3.5cm}{\begin{center}
      calculate\\ $C_{\ell_2}(b)=$\\[1mm]
      $\frac{\DS\mathrm{area}(\ell_2,b)}
            {\DS\mathrm{length}(\ell_2)}$\end{center}}}}
  \put( 5.75,7.5){\line(0,-1){0.5}}

  \put(11.8, 11.25){\makebox(2,0){forall $b$}}
  \put( 9.75, 8.5){\framebox(4,2){\parbox{4cm}{\begin{center}
      calculate\\ field line $b$\end{center}}}}
  \put(11.75, 8.5){\line(0,-1){1.5}}
  \put( 1.75, 7  ){\line(1, 0){10}}
  \put( 8,    7  ){\vector(0,-1){0.5}}

  \put( 3, 4.5){\framebox(10,2){\parbox{10cm}{\begin{center}
      determine fieldline\\$\widehat{b}(\ell_1,\ell_2)=
       \mathrm{argmin}(C_{\ell_1}(b)+C_{\ell_2}(b))$
                     \end{center}}}}
  \put( 8, 4.5){\vector(0,-1){1.5}}

  \put( 1.5,0.5){\framebox(13,2.5){\parbox{13cm}{\begin{center}
      for, e.g., $n_1>n_2$, determine the permutation $\pi$\\[-2mm]
      of $n_2$ elements from a set of $n_1$ which minimizes\\
      $ C_{    \ell_1 }(\widehat{b}(\ell_1,\pi(\ell_1)))
       +C_{\pi(\ell_1)}(\widehat{b}(\ell_1,\pi(\ell_1))))$
                     \end{center}}}}

\end{picture}
\caption{How does the magnetic field help us with stereoscopy?
The scheme is explained in detail in the text.
We take $\ell_1 \rightarrow \ell_2$ = $\pi(\ell_1)$ as the
association between loops in image 1 and 2.}
\label{fig0}
\end{figure}
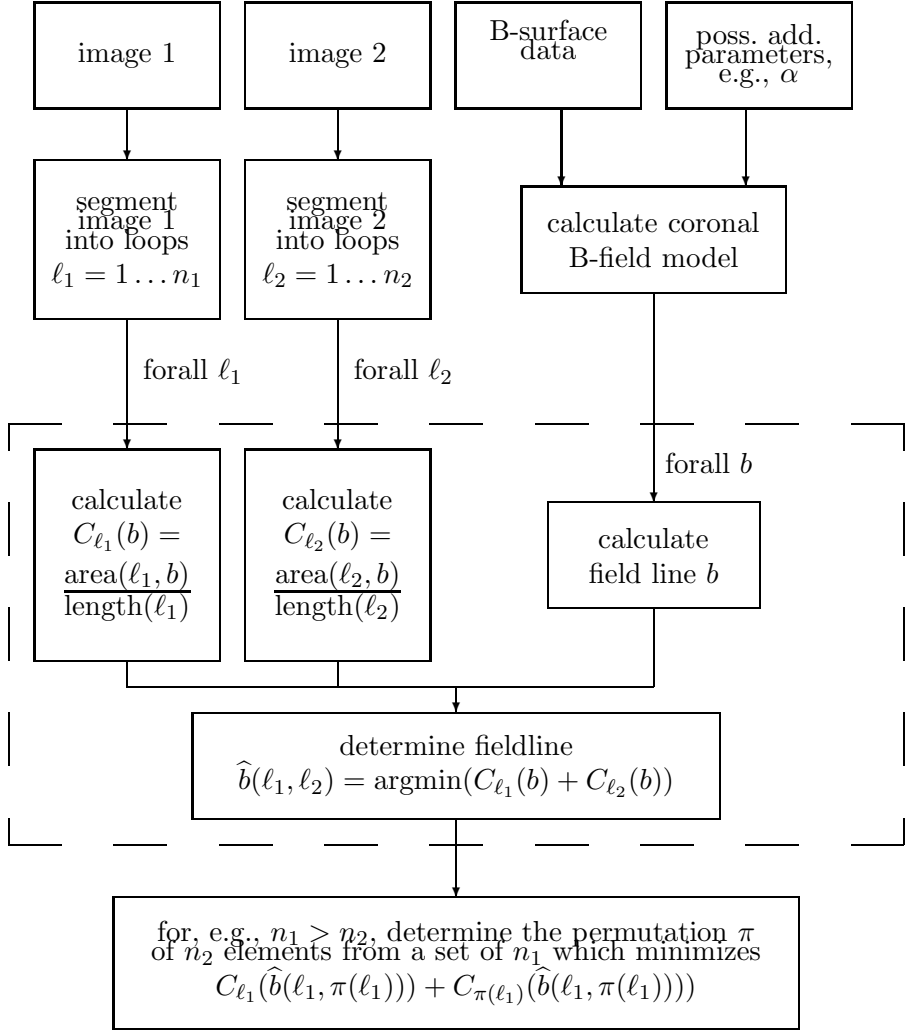
\begin{figure}
\mbox{
\includegraphics[bb=40 20 385 320,clip,height=5cm,width=8cm]{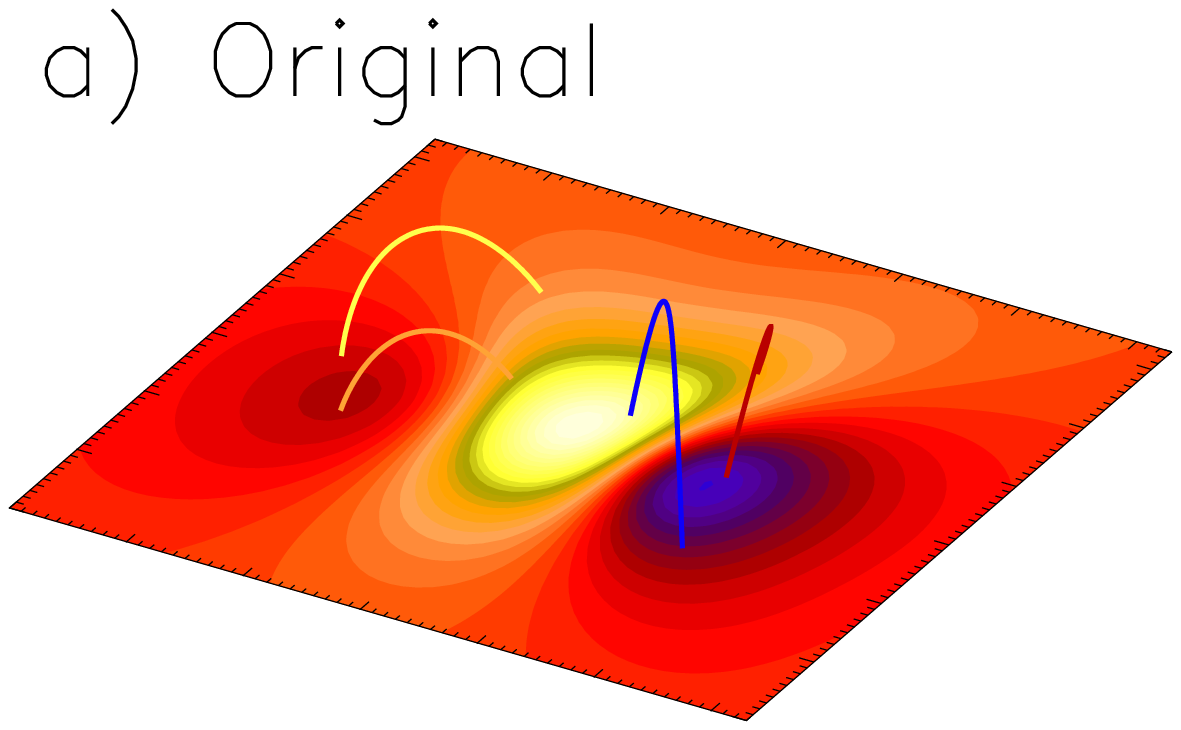}
\includegraphics[bb=40 20 385 320,clip,height=5cm,width=8cm]{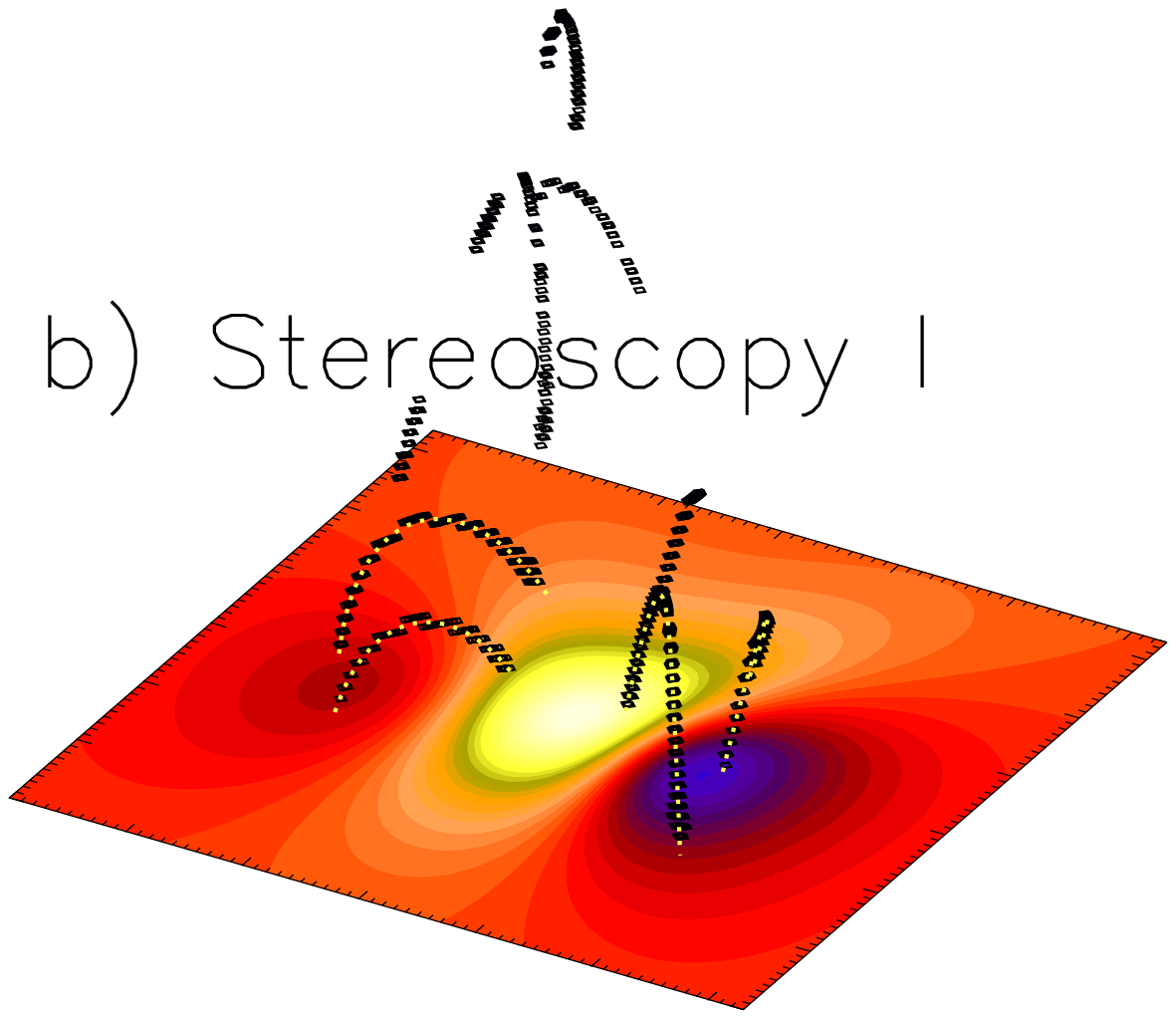}
}
\mbox{
\includegraphics[bb=40 20 385 320,clip,height=5cm,width=8cm]{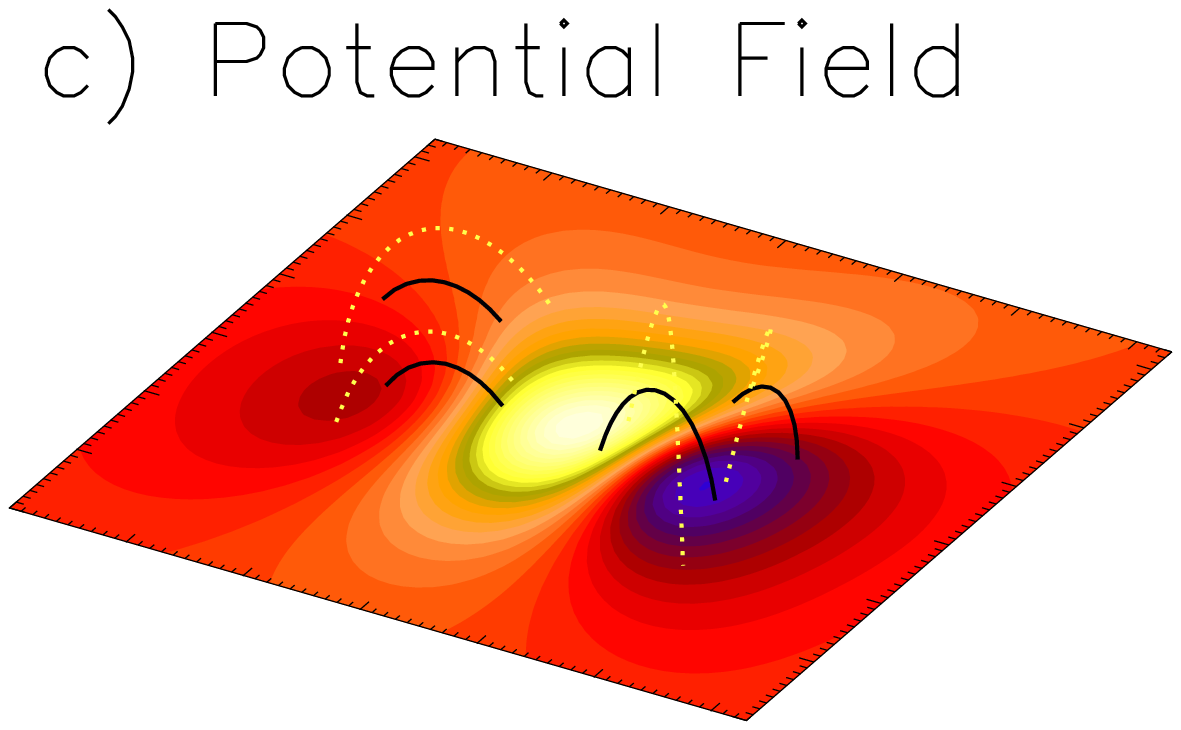}
\includegraphics[bb=40 20 385 320,clip,height=5cm,width=8cm]{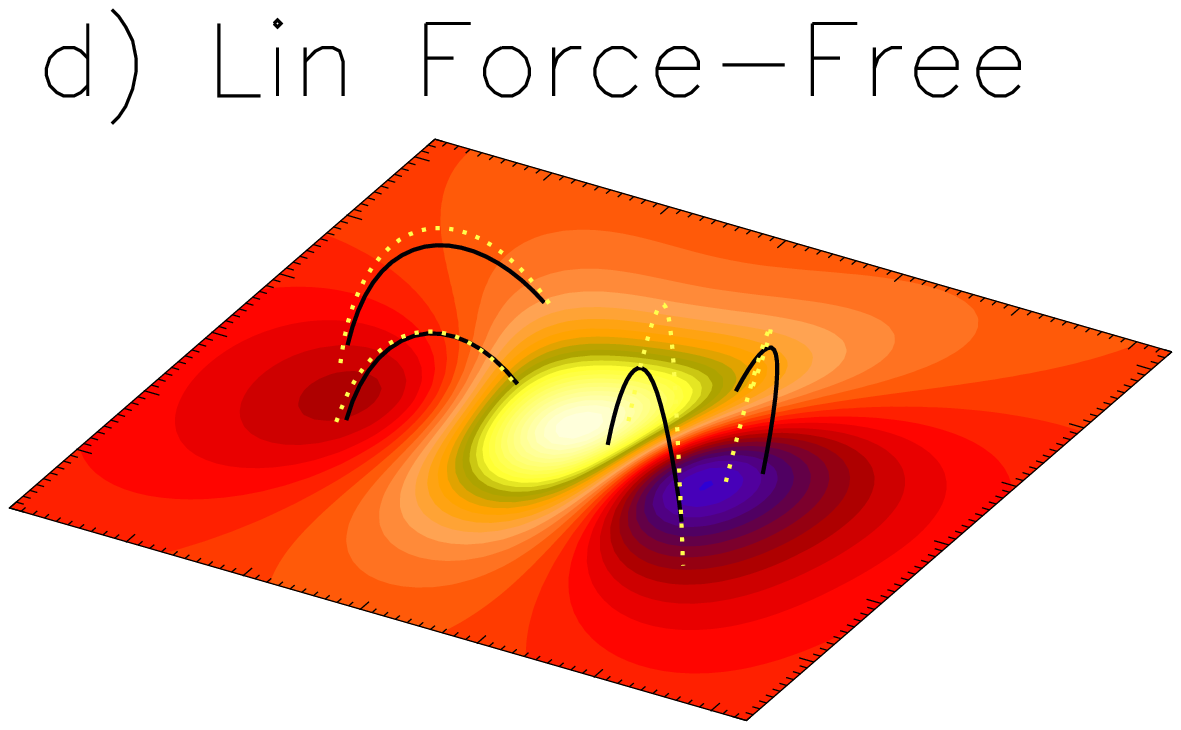}
}
\mbox{
\includegraphics[bb=40 20 385 320,clip,height=5cm,width=8cm]{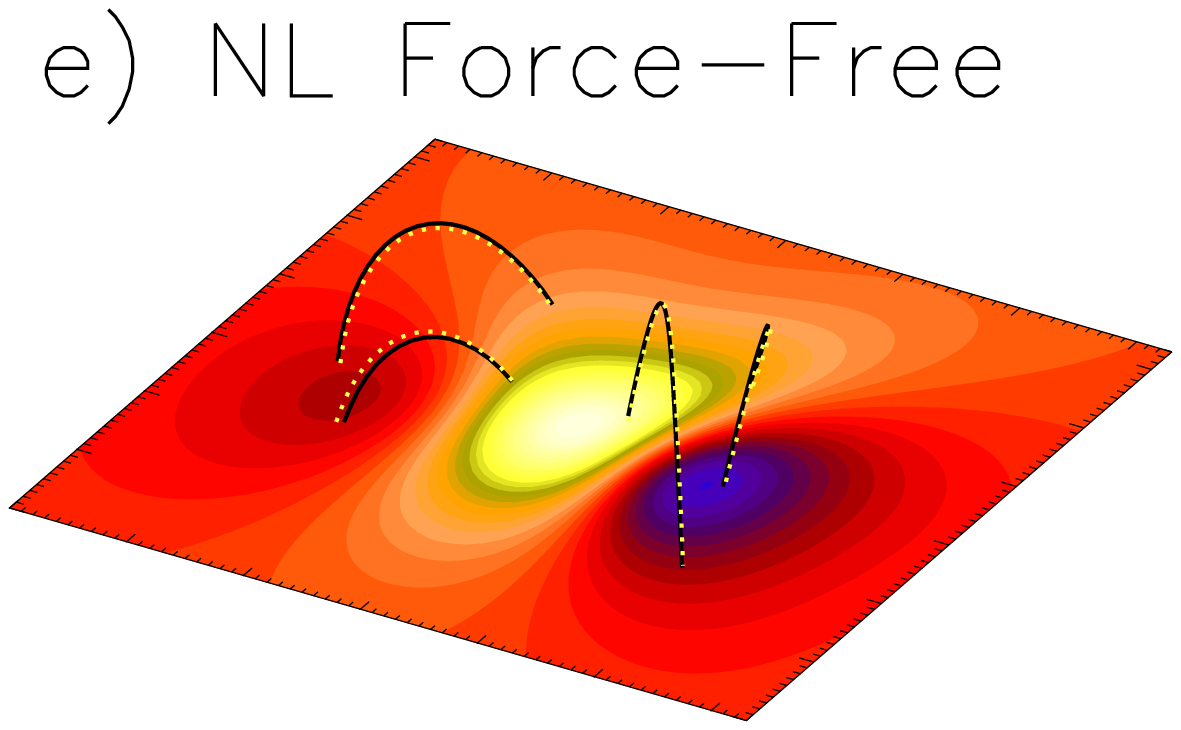}
\includegraphics[bb=40 20 385 320,clip,height=5cm,width=8cm]{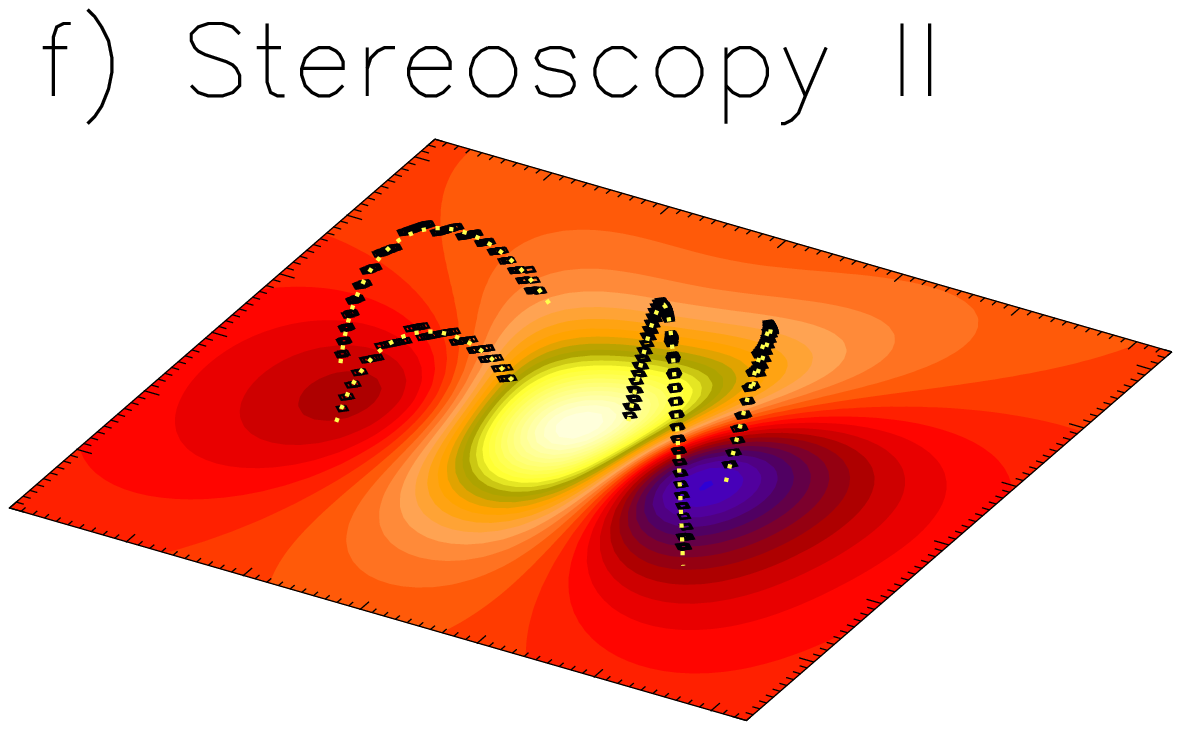}
}
 \caption{a) Model active region. We use the model
 developed by Low and Lou (1990) with the parameters $l=0.5$ and $\Phi=1.4$.
 We show four loops
({\it 1- blue, 2-red, 3-orange, 4-yellow}). The other panels show different method
of reconstruction this loops (from the two images in Figure \ref{fig2}). The
reconstruction is shown in {\it black} and for comparison the original loops dotted
in {\it yellow}.}
 \label{fig1}
\end{figure}
\begin{figure}
\mbox{
\includegraphics[bb=0 0 420 420,clip,height=8cm,width=8cm]{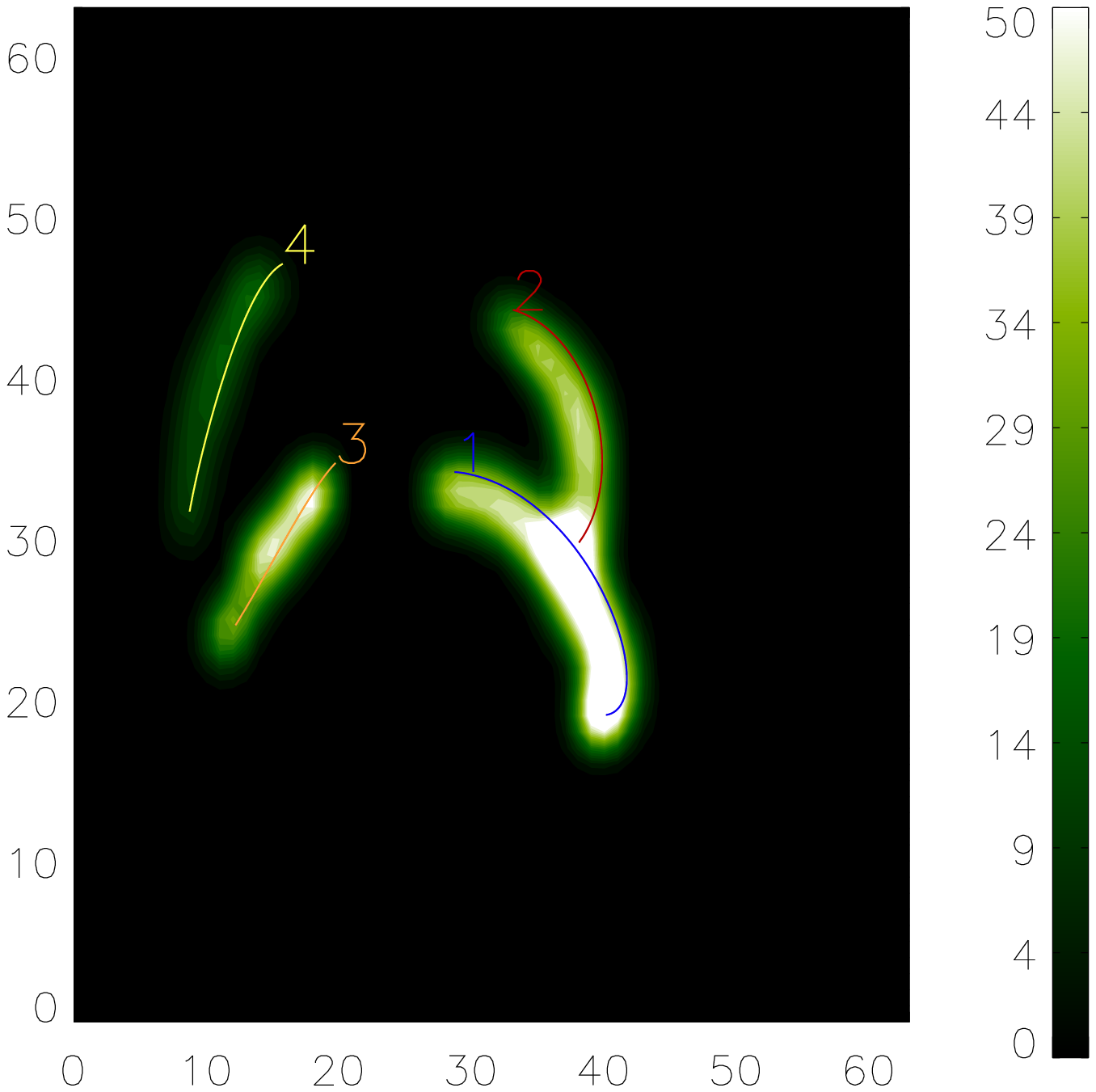}
\includegraphics[bb=0 0 420 420,clip,height=8cm,width=8cm]{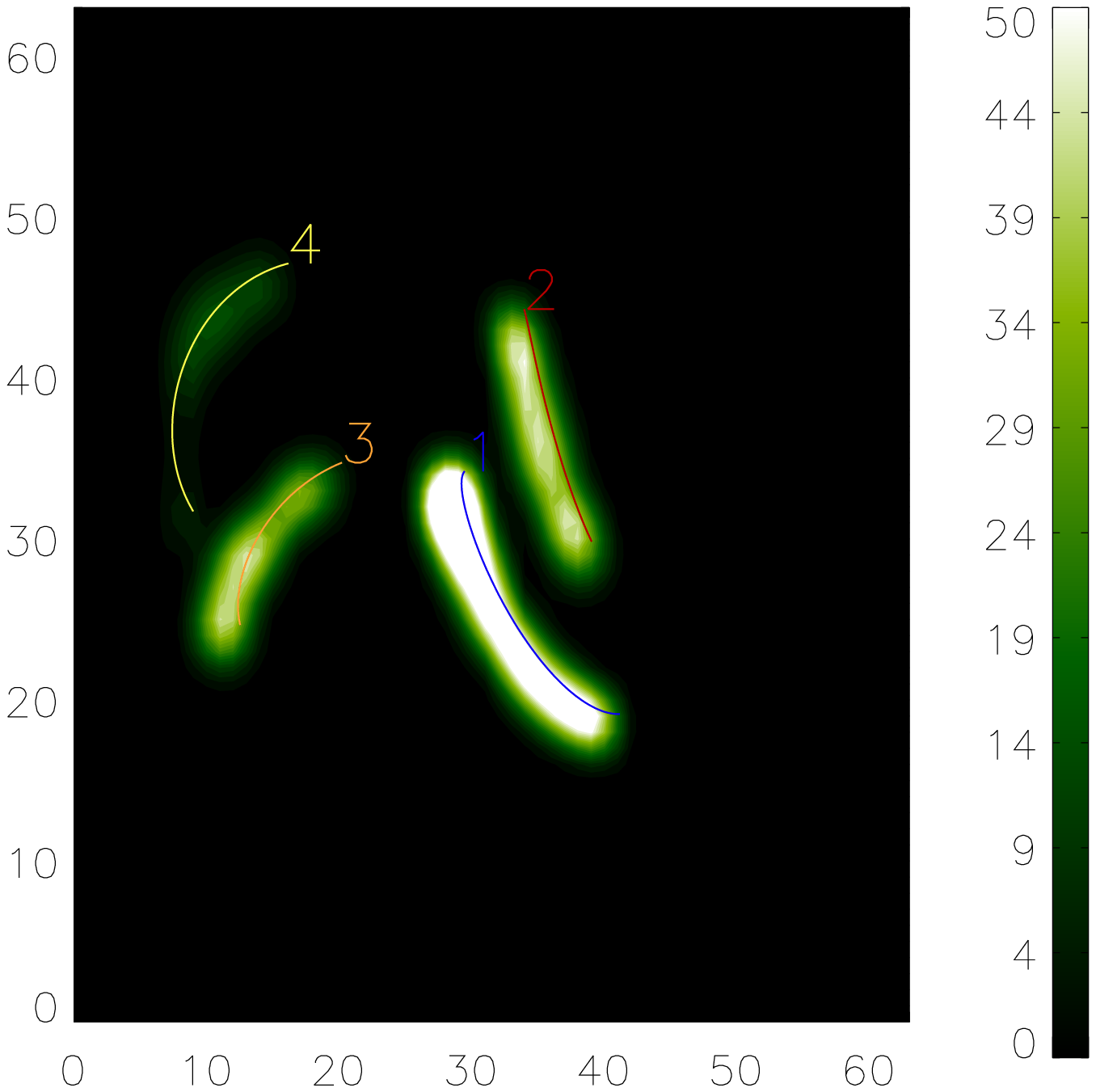}
}
 \caption{Artificial STEREO-images (STEREO-1 in the left and
 STEREO-2 in right panel) with an angle between the spacecraft
 of $56^o$. The coloured lines show the projections of the original
 3D loops ({\it 1-blue, 2-red, 3-orange, 4-yellow}).}
 \label{fig2}
\end{figure}
\begin{figure}
\mbox{
\includegraphics[bb=40 10 400 430,clip,height=8cm,width=8cm]{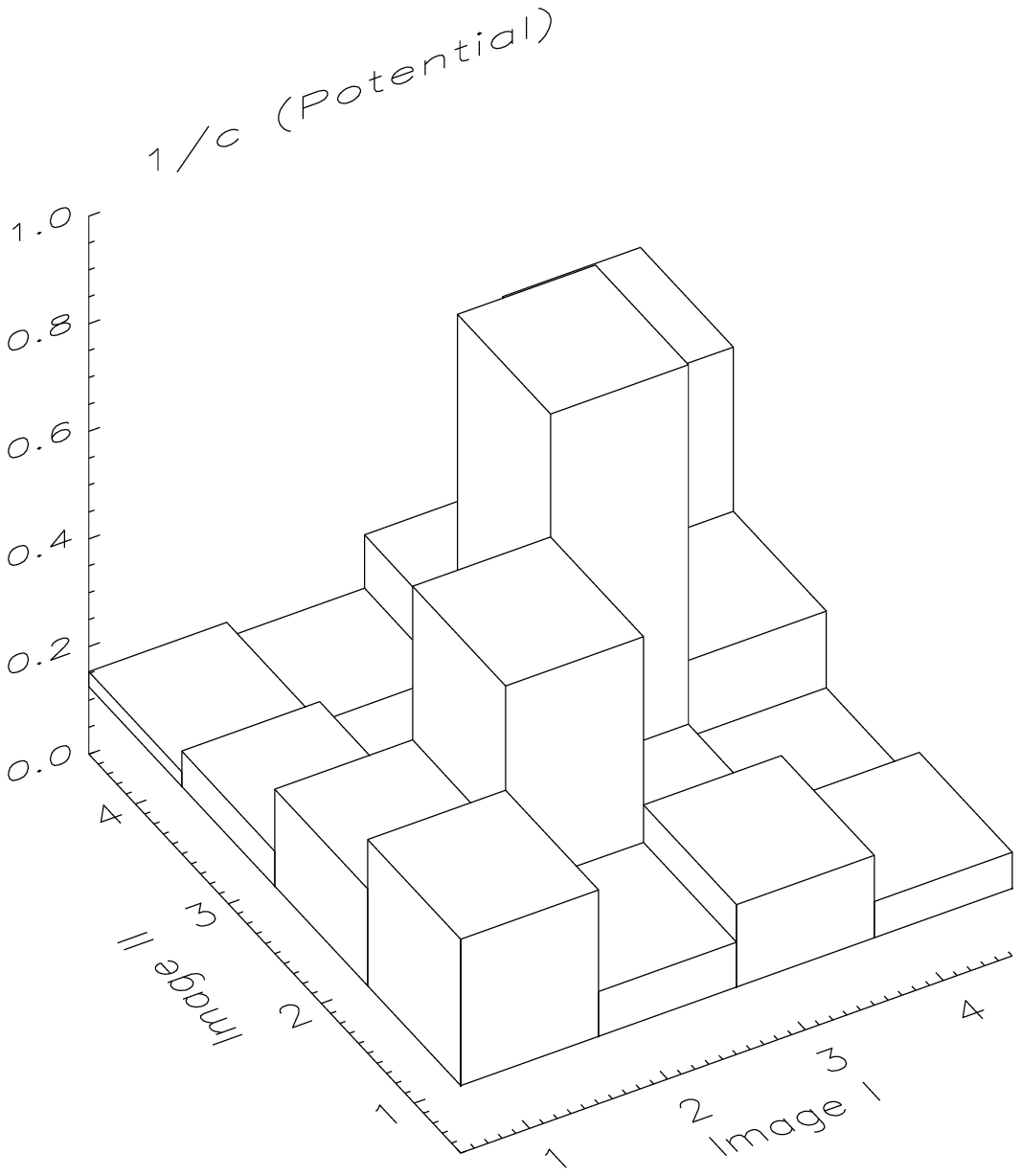}
\includegraphics[bb=40 10 400 430,clip,height=8cm,width=8cm]{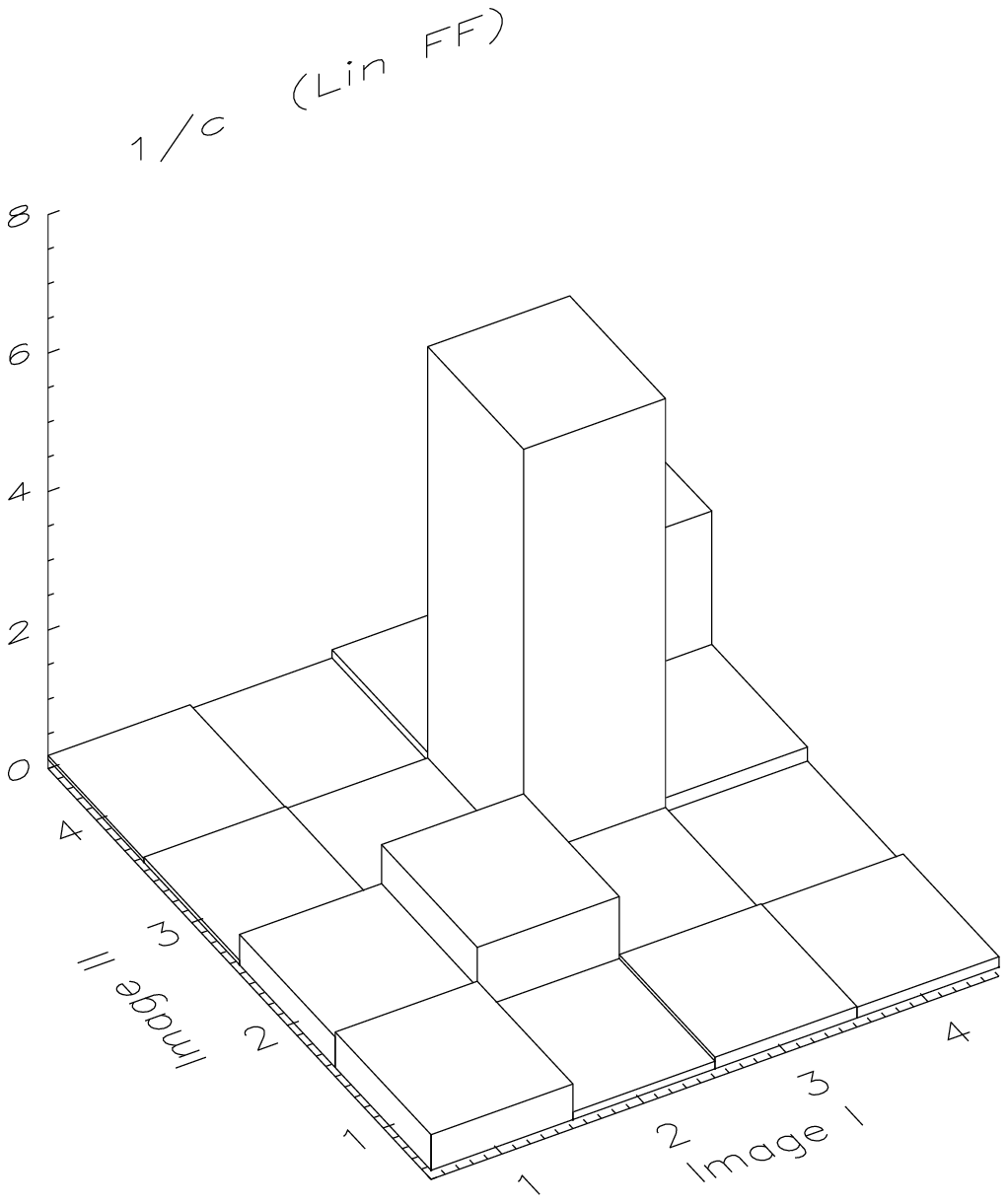}}
\mbox{
\includegraphics[bb=40 10 450 430,clip,height=8cm,width=8cm]{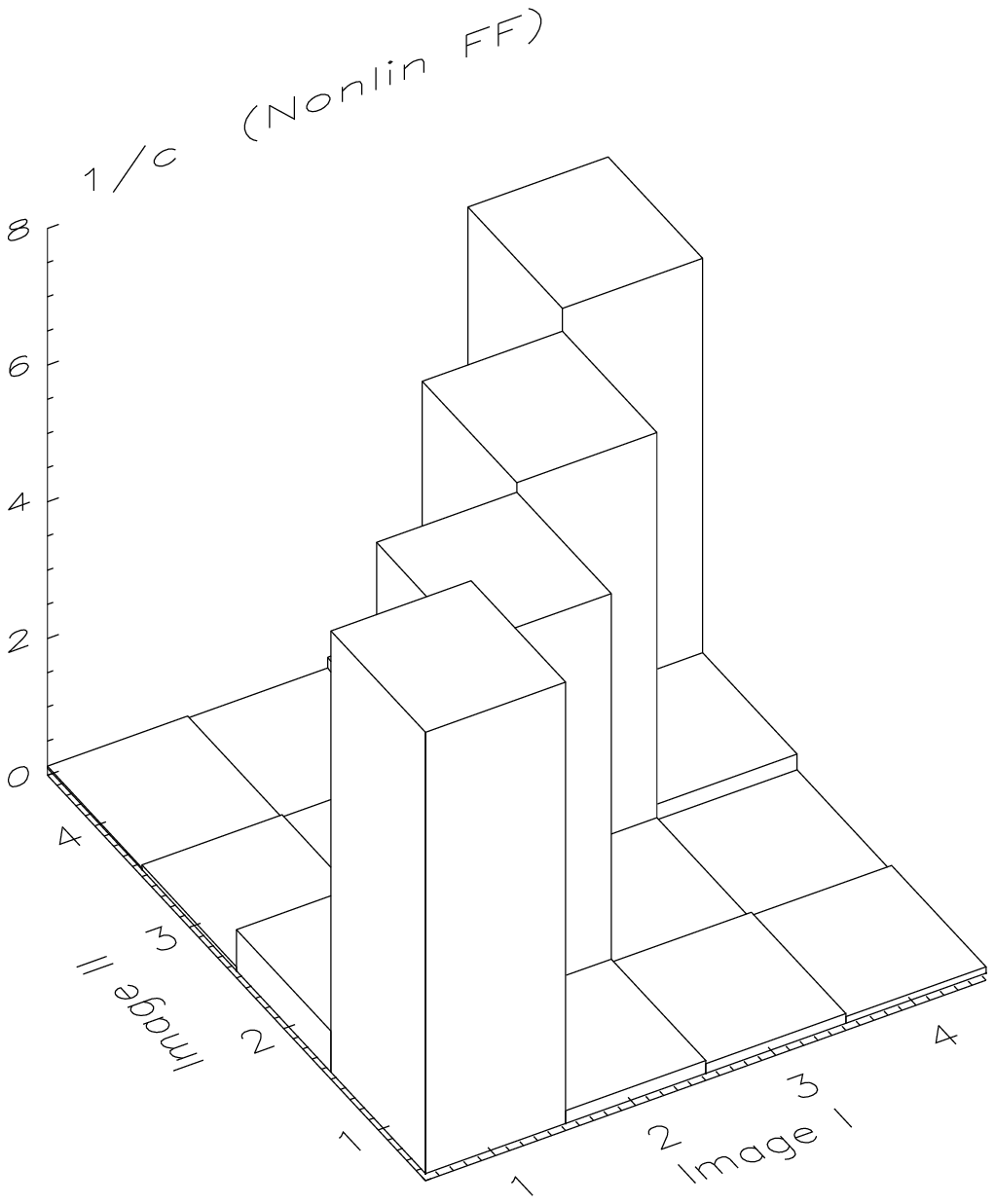}
\includegraphics[bb=40 10 450 430,clip,height=8cm,width=8cm]{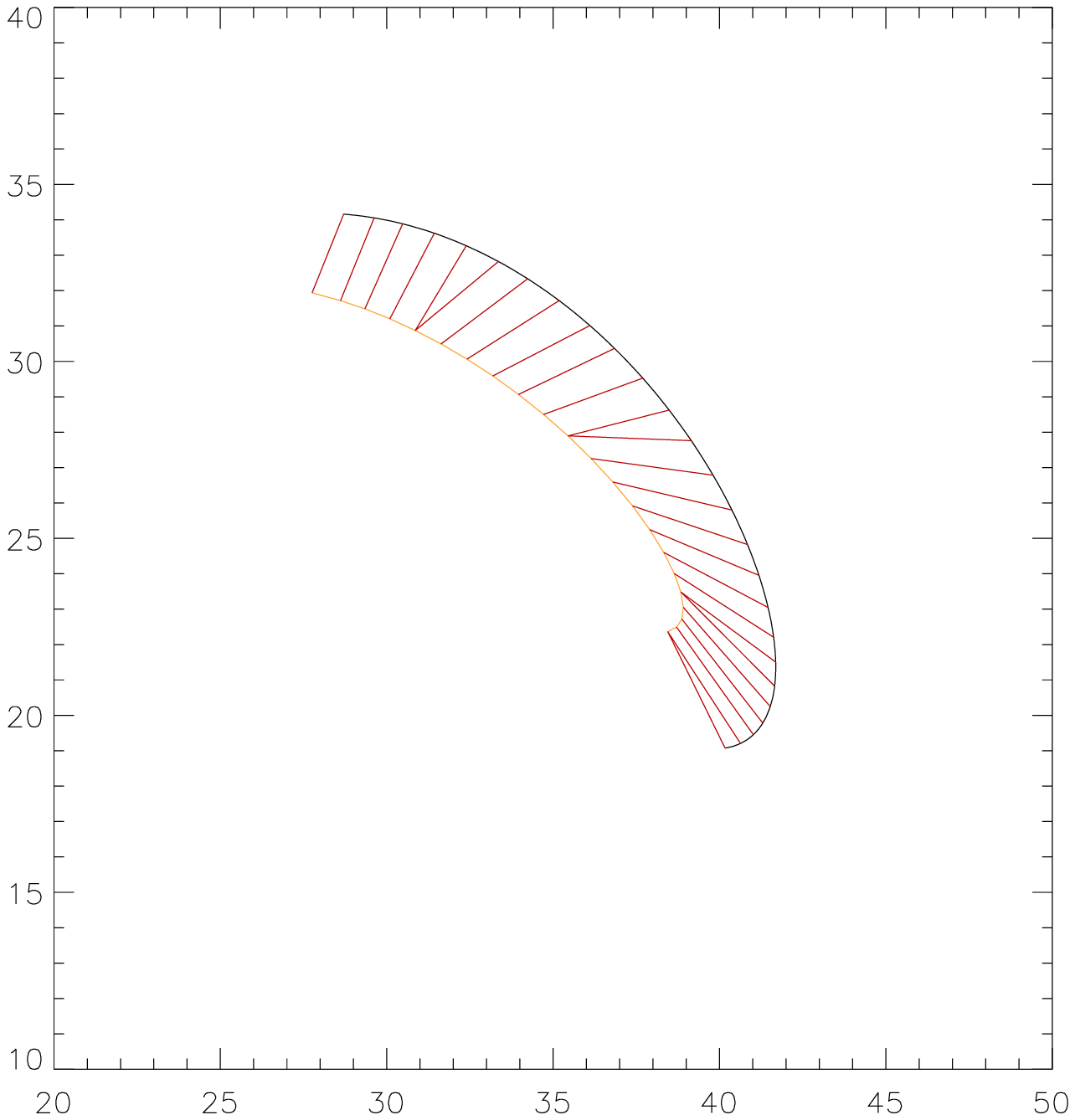}
} \caption{Loop association with different coronal magnetic field models. The {\it
upper panels} and the {\it lower left panel} show the matrix $1/C$ which associates
each loop in {\it image 1} with each loop in {\it image 2}. We plot $1/C$ instead of
$C$ because the best association of loops corresponds to maxima here, which are
better visible than minima in the stacked histogram-style plots. The {\it lower
right panel} illustrates the area between a loop (from one STEREO-image) and a
magnetic field line projection. $C$ is defined as the area divided by the length of
the loop.} \label{fig3}
\end{figure}

\section{How does the magnetic field help us with stereoscopy?}
The scheme in Figure \ref{fig0} outlines how information regarding the coronal
magnetic field can be used to improve the stereoscopic reconstruction. A key
question is to associate coronal features, e.g., loops in image 1 (from the STEREO-1
spacecraft) with features in image 2 (observed from STEREO-2). In the following we
specify some details of the scheme.
\begin{enumerate}
\item A first step is to segment the 2D images into loops. This is by no means a
trivial process and several methods have been proposed, e.g., the brightness
gradient method, the oriented-connectivity method, magnetic field extrapolation and
curvature radius constraints and the use of multiple temperature filters. An
overview about current developments for this step is given in
\cite{aschwanden05,lee:etal06}.

\item The next step requires a suitable magnetic field model. As the coronal
magnetic field can usually not be measured directly, we extrapolate it from
photospheric magnetic field measurements (from vector or line-of-sight
magnetographs). The coronal magnetic field model may also contain additional
parameters, e.g., the linear force-free parameter $\alpha$. An overview about
coronal magnetic field models is given in section \ref{magnetic_modelling}. We use a
fourth order magnetic field line tracer to compute 3D magnetic field lines from the
3D magnetic field. The starting points are chosen randomly, magnetic flux weighted,
on the photosphere and in the current implementation only closed magnetic loops
(both footpoints are on the photosphere) are stored because the main application is
to identify closed coronal loops, see also \cite{wiegelmann:etal05b}. Open field
lines are expected not to be visible anyway. As a result we get space filling
magnetic field lines in 3D. A basic assumption is that the emitting plasma loops
also outline the coronal magnetic field. This is a consequence of the high
conductivity of the coronal plasma. The emissivity gradient along the magnetic field
is much smaller than in the perpendicular direction.

\item The 3D magnetic field lines are projected onto both images from the two
STEREO-spacecraft. For every projected field line we measure how well it agrees with
one of the loops identified in step one. As a measure of this agreement we take the
area between each loop and the projections of the magnetic field line in each image
(see lower right panel in Figure \ref{fig3}) normalized by the length of the
respective loop. The measure $C=C_{\ell_i}(b)$ then corresponds to the average
distance of the loop and the projection of the 3D magnetic field line. Perfect
agreement corresponds to $C=0$. If $n_b$ is the number of field lines and
$n_{\ell_i}$ the number of loops in image $i$ ($i=1$ or $i=2$) we get a matrix of
the dimension $n_b \times n_{\ell_1} $ for image one and another $n_b \times
n_{\ell_2}$ matrix for image two, which contain the corresponding values $C_{l1}$
and $C_{l2}$. We here give C in units of a pixel size to have a clear relation to
the image resolution.
\item The next step is to associate the loops in both images which each other. The
aim is to find for each pair of loops $l_1$ and $l_2$ a magnetic field line $b$ for
which the summed measures $C_{l1}(b)$ and $C_{l2}(b)$ are minimal. As a result of
this step we get a $n_{\ell_1} \times n_{\ell_2}$ matrix which contains the
arithmetic mean of $C_{l1l2}(b)=(C_{l1}(b)+C_{l2}(b))/2$ from both images for all
possible association between image 1 and image 2. $C_{l1}+C_{l2}$ can be calculated
for each magnetic field line. The optimal magnetic field line $b$ corresponds to the
minimum of $C=C_{l1l2}$ in $b$. Again, $C=0$ corresponds to perfect agreement.
Figure \ref{fig3} contains a stacked histogram-style plot for this matrix ($4 \times
4$ matrix for the four loops features shown in Figure \ref{fig2} from two
viewpoints. We preferred to plot $1/C$ instead of $C$ for a better visualization.
Here high values of $1/C$ correspond to a good agreement, e.g., a value of $1/C=5$
means that the projection of the field line and the loop are in average only one
fifth of a pixel apart.) The absolute values $C$ certainly depend on the chosen
magnetic field model and partly also on the number of field lines plotted in step 2.

For a good magnetic field model one has a clear association of features in image 1
with features in image 2 after this step already. This is certainly the case for the
non-linear force-free model shown in the lower left panel of Figure \ref{fig3}. The
method correctly associates loop 1 of image 1 with loop 1 of image 2 etc. with
values of $C < 1/5 {\rm pixel}$. $\; (1-1, \; C=0.15), \; (2-2, \; C=0.18), \; (3-3,
\; C=0.17), \; (4-4, \; C=0.16)$, which gives an average measure of $C= 0.17 \pm
0.01$.

As a consequence the 3D magnetic field lines are already an excellent proxy for the
3D loop and an explicit geometric stereoscopic reconstruction is not necessary
anymore. Figure \ref{fig1} e) shows the four identified magnetic field lines in
black and the original model loops dotted in yellow. The picture shows an agreement
of original and reconstruction within plotting precision.

The loop association with help of potential and linear force-free fields (upper left
and right panel in Figure \ref{fig3}) is not as good as for the non-linear
force-free case. The linear force-free model associates the loops  in both images
with a distance of $(1-1, \; C=1.54), \; (2-2, \; C=0.89), \; (3-3, \; C=0.16) \;
(4-4, \; C=0.44)$ or in average $C=0.76 \pm 0.60$. All loops are associated
correctly. The linear force-free model provides us also the optimal values of
$\alpha$ for each loop, which are $\alpha \, L= -4$ for loop 1 and 2 and $\alpha \,
L= +4$ for loop 3 and 4. Different values of $\alpha$ on different loops are a
contradiction to the assumption of a linear force-free model, which requires a
single global value of $\alpha$. So the loop association method tells us also
whether a linear force-free field is a fair approximation for the coronal magnetic
field ($\alpha$ is identical or almost identical for all loops) or not. In this
example it is not. Nevertheless the optimal linear-force free 3D magnetic field
lines are a proxy for the 3D plasma loop. The proxy is the better the smaller the
correspondent values of $C$. Figure \ref{fig1} d) shows the four proxy field lines
for the linear force-free case. In accordance with the values of the matrix (upper
right panel of Figure \ref{fig3}) we get a very good agreement with the original for
loop 3, some small deviations for loop 4 and larger deviations for loop 1 and 2.

The potential field (lower left panel of Figure \ref{fig3}) produces quite high
values of $C$ and the distances between projected field lines and loops are mostly
larger than $1 \, {\rm pixel}$ or $(1/C < 1)$. The potential magnetic loops are not
a good approximation for the reconstructed plasma loops due to the quite high values
of $C$ and as visible in Figure \ref{fig1} panel c). The reconstruction (black) and
original (yellow) are obviously far apart.

The potential field yet produces a correct association of the loops.
$(1-1, \; C=2.52), \; (2-2, \; C=1.69), \; (3-3, \; C=1.22), \; (4-4, \; C=1.74)$ or
in average $C=1.79 \pm 0.54$. Let us also note the lowest incorrect associations here
$(1-2, \; C=3.27), \; (3-1, \; C=3.59), \; (4-3, \; C=3.70), \; (3-4, \; C=4.45)$, which
are only slightly larger than the correct associations. If $C>1$ the correct
loop association might not be absolutely clear after this step.
In such a situation
a further step is necessary to
associate all features in both images with each other.

\item If step 4 does not provide a clear association of features in the two images
we need to undertake a further step. Here we check which combination of association
of features will give the lowest values of $C$ (best agreement). In principle one
would restrict this analysis to critical loops, which cannot be associated clearly
because for $n$ features there are ${\rm factorial \, of}(n)$ possible combinations.
For our test example $(n=4)$ we computed all ${\rm factorial \, of}(4)=24$
possibilities \footnote{If the number of loops in the two STEREO-images are
different, say $n_a < n_b$, we get $\frac{n_b!}{(n_b-n_a)!}$ possible
permutations.}.

For every
combination we compute the mean of the $4$ C-values. The combination with the lowest
value of $C$ is the most likely one. With all three magnetic field models
(potential, linear force-free, non-linear force-free) we get the correct association
$1-1, \; 2-2, \; 3-3, \; 4-4$ as the most likely one. The correspondent values of
$C$ are presented in the upper part of table \ref{table1} (Example 1).
\begin{table}[h]
\caption{Here we show for two examples how good the loops in both STEREO-images
can be associated with each other for different magnetic field models. As we associate four
loops in both images, there is a total number of $24$ possible combinations. Here we shown only the
best (lowest value of $C$) three combinations.}
\label{table1}
\begin{tabular}{llll}
B-field model & $C_{\rm best}$ & $C_{\rm 2. best}$& $C_{\rm 3. best}$ \\
\hline
Example 1 && \\
NonLin FF     & $0.17$ &$1.76$ &$2.02$ \\
Lin FF        & $0.76$ &$2.10$ &$2.56$ \\
Potential     & $1.79$ &$2.76$ &$3.06$ \\
\hline
Example 2 && \\
NonLin FF     & $0.16$ &$0.89$ &$1.31$ \\
Lin FF        & $0.87$ &$1.52$ &$1.96$ \\
Potential     & $1.54$ &$1.54$ &$1.92$ \\
\end{tabular}
\end{table}
 All three magnetic field models give the lowest value of $C$ to the correct
 loop association.  The best magnetic
 field model (nonlinear force-free) gives also the clearest answer regarding the optimal loop
 association. C for the optimal combination is a factor of $10$ lower than
 the second best combination. For linear force-free fields the best combination
 is still a factor of almost $3$ better than the second best. For potential field
 the second best combination is only about a factor of $1.5$ higher than the best one. This
 might still be enough to get some evidence about the correct association of
 loops in the two STEREO images. For the application to observational data
 one might include a threshold, e.g., consider only loops as clearly associated
 which each other when (the pairwise) value of C is lower than a certain threshold (say
 e.g.,
 $1 \, {\rm pixel}$). Larger values of $C$ might be in particular problematic if
 the loops or features are closer together then here. We apply our method
 to such a case in section \ref{example2}.
\item As the last step we do a geometric stereoscopic reconstruction, similar as
 described in section \ref{Geometric_Stereoscopy}, but now with the knowledge which
 pairs of loops in the two images have to be used for the
 stereoscopic reconstruction. This knowledge removes already most of the ambiguities
 (visible as {\it ghosts} in Figure \ref{fig1} b).

 But even if we have identified pairs of
 loops in the two images this does not mean that we can identify
 each pixel along two associated curves in both images with each other. Consequently
 there can still be multiple points of intersection (and thus not a unique solution of the
 3D reconstruction). Again, the magnetic field proxy (but now in 3D)
 can help here to resolve the ambiguity. From multiple point of intersection we chose that point,
 which has the closest distance to the magnetic field proxy. Even if we use a less
 than optimum magnetic field model we get
 a very good geometric stereoscopic reconstruction as seen in Figure \ref{fig1} f) where a
 potential field has been used to remove the ambiguities of multiple intersection points.
 All stereoscopic reconstructed points (black) coincide with points of the original
 loops (yellow dotted). The {\it ghost points} (Figure \ref{fig1} b) have vanished.
  One can see that the combination of
 geometric stereoscopy and magnetic modelling in Figure \ref{fig1} f) is much more
 powerful than pure stereoscopy (Figure \ref{fig1} b) or potential field magnetic modelling
 (Figure \ref{fig1} c) alone.
 We call this combination of geometric (or ordinary) stereoscopy with magnetic
 modelling {\it magnetic stereoscopy}.
\end{enumerate}
\subsection{Example 2}
\label{example2}
\begin{figure}
\mbox{
\includegraphics[bb=0 0 420 420,clip,height=6cm,width=6cm]{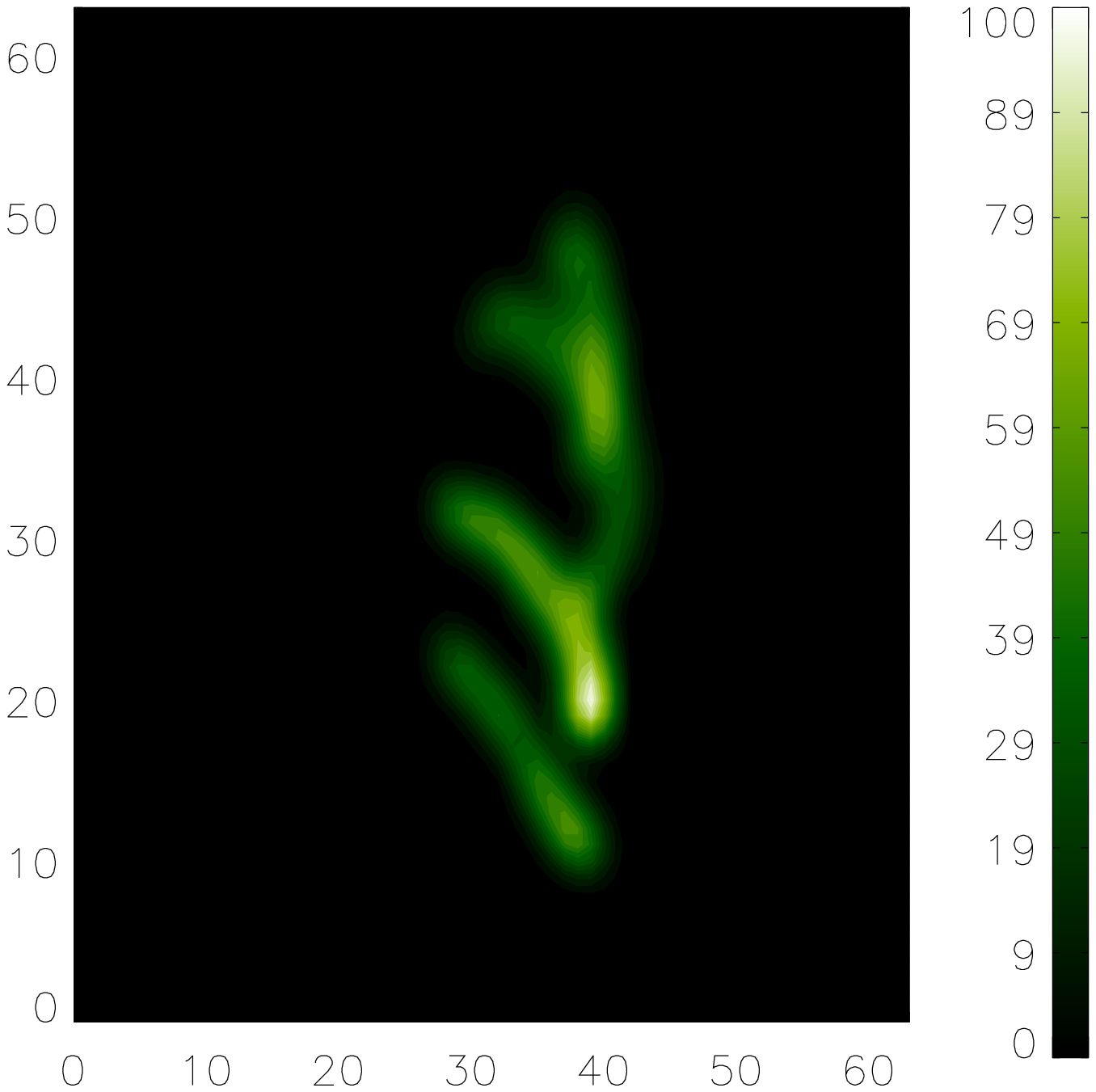}
\includegraphics[bb=0 0 420 420,clip,height=6cm,width=6cm]{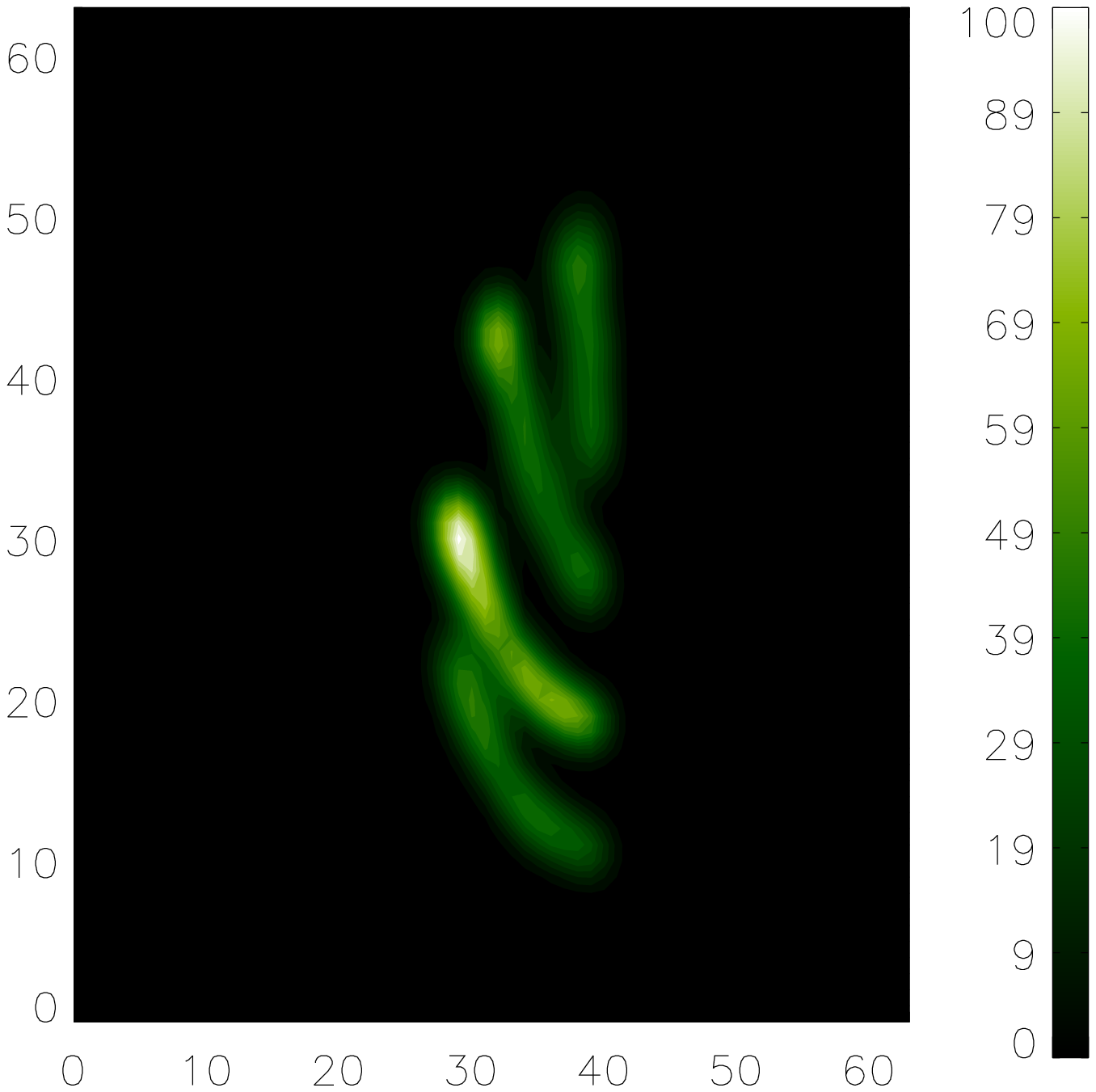}
}
\mbox{
\includegraphics[bb=160 90 260 340,clip,height=9cm,width=6cm]{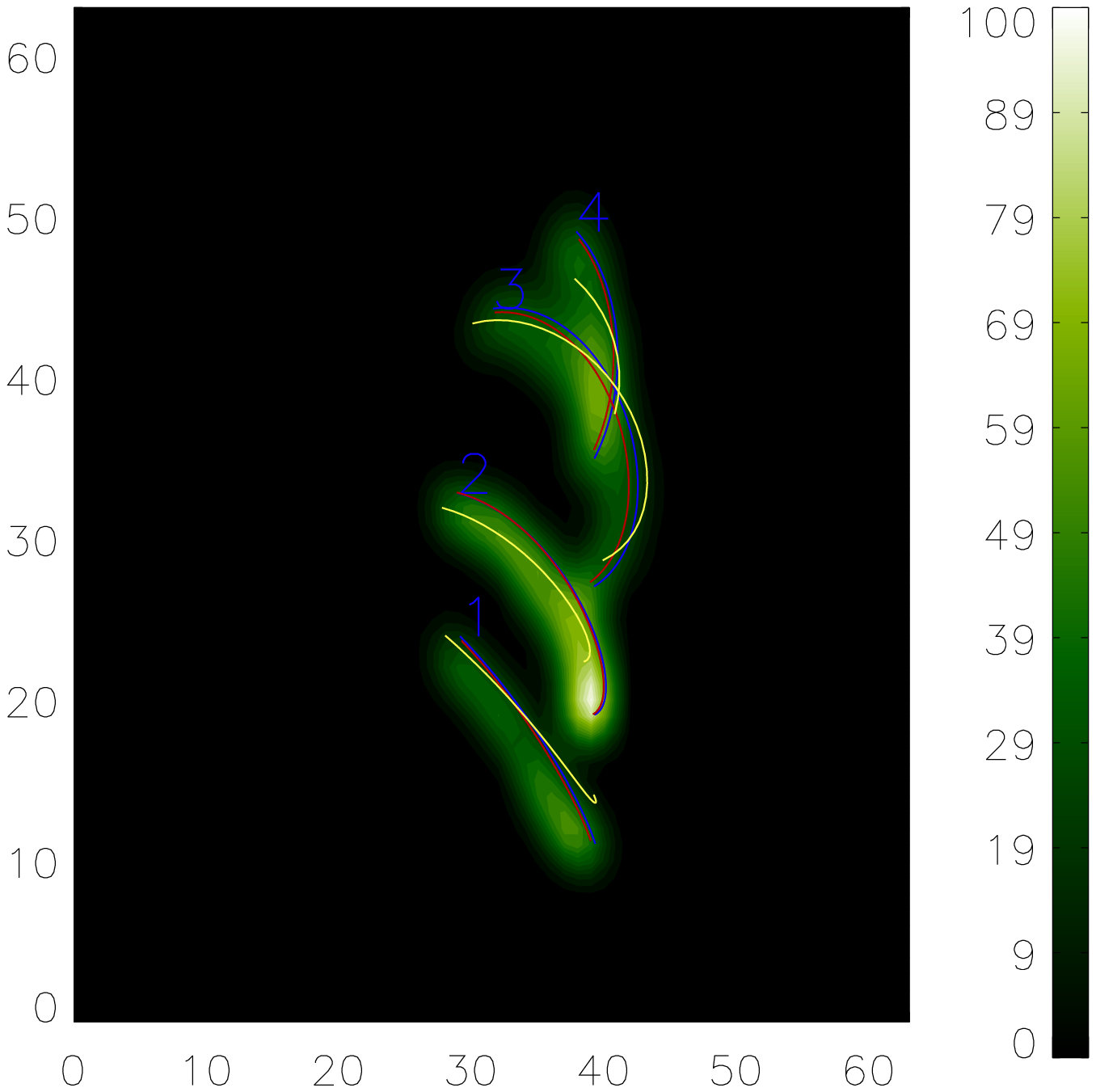}
\includegraphics[bb=160 90 260 340,clip,height=9cm,width=6cm]{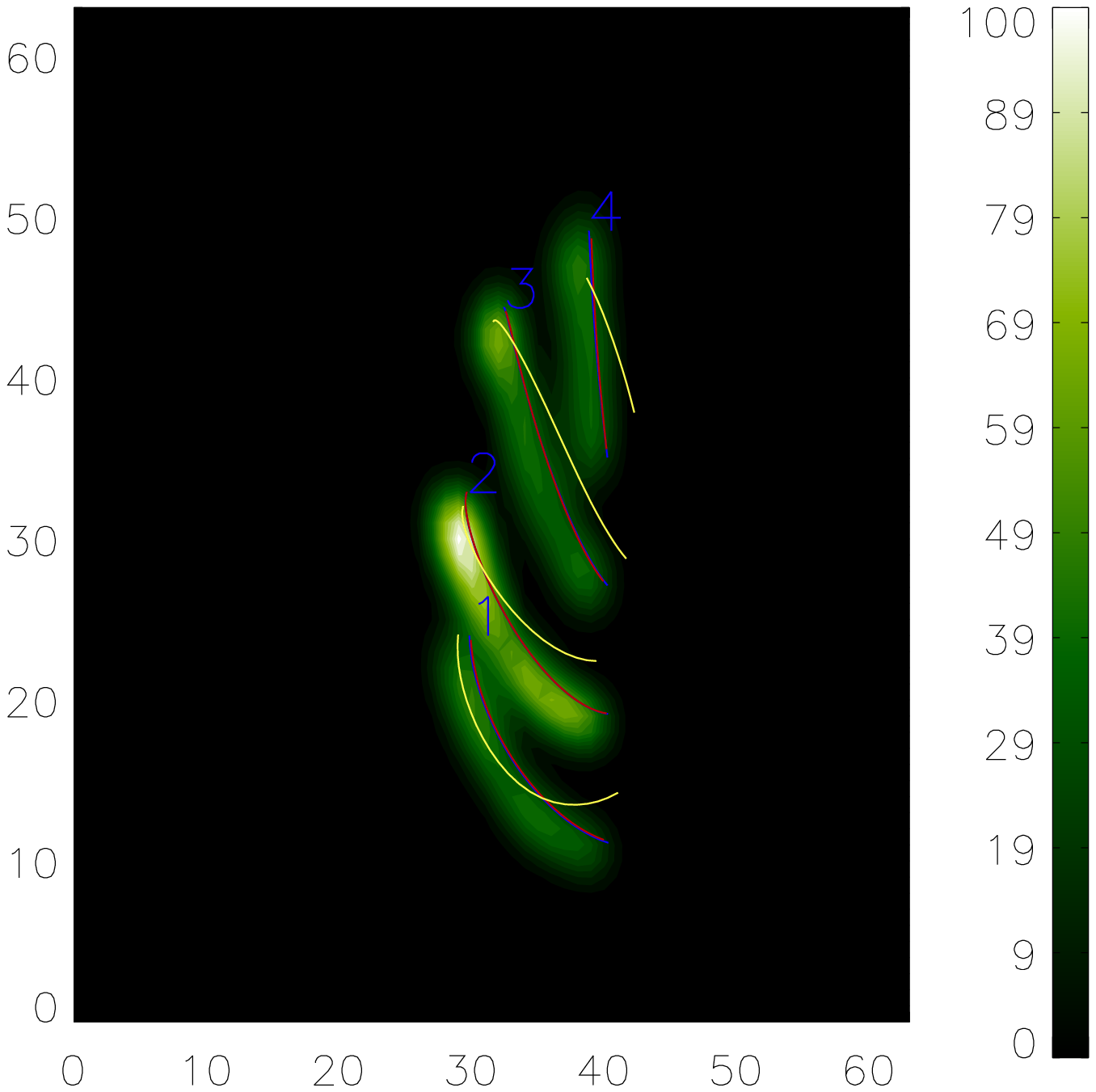}
}
\mbox{
\includegraphics[bb=190 50 300 190,clip,height=6cm,width=6cm]{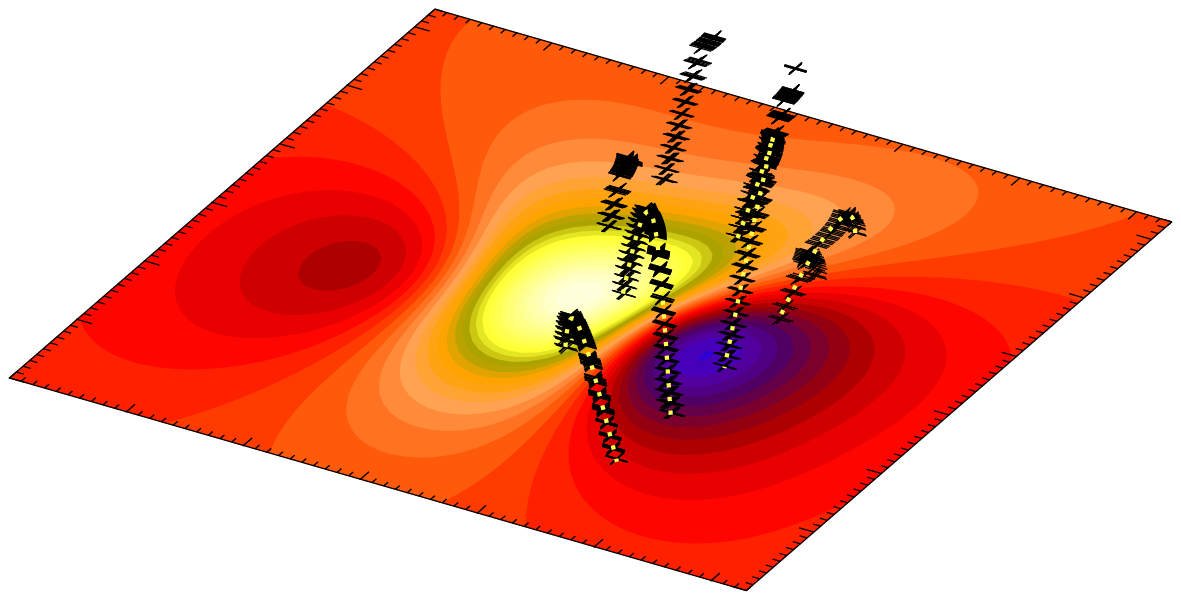}
\includegraphics[bb=190 50 300 190,clip,height=6cm,width=6cm]{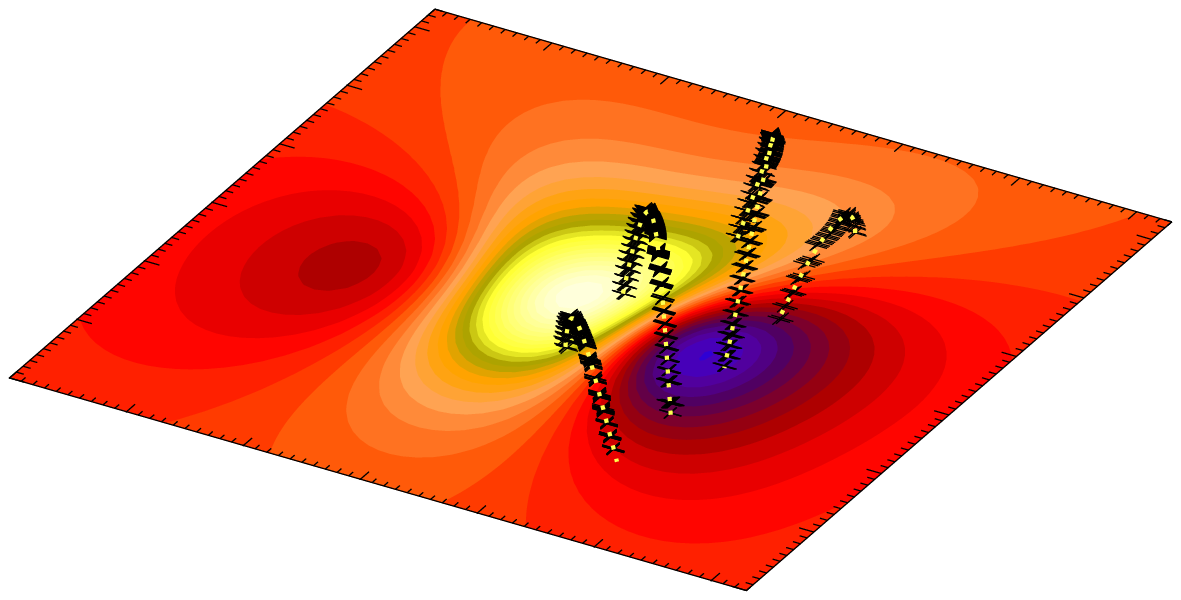}
}
 \caption{Example 2.:
 Left panels: STEREO-1, Right panels: STEREO-2.
 Top: Artificial EUV-images
 Center: EUV-images and projections of original loops ({\it blue}), nonlinear
 force-free loops ({\it red}) and linear force-free loops with $\alpha L=-4$
 ({\it yellow}).
 Bottom left:  Pure geometric stereoscopy.
 Bottom right: Magnetic stereoscopy using a linear force-free magnetic field model.}
 \label{example2fig}
\end{figure}
As a second example we apply our method to a set of loops that are closer together.
The upper panel of Figure \ref{example2fig} shows two artificial EUV-images from
different viewpoints (STEREO-1 and STEREO-2 in the right and left panel,
respectively). The center column shows the structures enlarged and we overplot the
image with different loops. The original loops are shown in {\it blue} and the
projection of the nonlinear force-free field lines in {\it red}. As one can see the
difference is very small. For loop-1 and 2 the {\it red} (nonlinear force-free) and
{\it blue} (original) lines can hardly be distinguished in the right hand (STEREO-2)
image. The projected images of the linear force-free field ({\it yellow}) are
somewhat apart from the original.

Table \ref{table1} lower part (Example 2) shows
the quantitative measures (average value of $C$ as explained in the last section)
for the optimal loop association as well as for the second and third best combination.
For a nonlinear force-free model the average distance is (as in example 1) less than
$1/5$ of a pixel and this model gives the clearest answer regarding the correct
association of loops as the second best loop association is a factor of $5.5$ worse
then the correct one.

The linear force-free model provides a considerably higher value for the optimum
combination of $C=0.87$, which is, however, almost a factor of two $(1.74)$ better
than the second best combination. Here (because we know the exact solution) we can
also confirm that the optimum (lowest value of C) combination is the correct one.
For real STEREO-data (where we do not know the correct solution of course) one has
to trust merely on the values of C alone. The higher values of C for the linear
force-free model tell us that this model is worse than the nonlinear force-free one.
Here the linear force-free parameter $(\alpha L=-4)$ was the same for all loops.

Similar as in example 1 the potential magnetic field gives the worst results.
And here this model does not provide a clear association of the correct loops
in both images to each other. Two combination (including the correct one) provide
the same average value of C. We therefore conclude that a potential field is a
too simple approximation.

In the left hand bottom panel of Figure \ref{example2fig} we show a pure geometric
loop reconstruction ({\it black}) and the original loops in {\it yellow}. The
reconstruction contains ghost features. In the bottom right hand panel we use
magnetic stereoscopy, similar as described for the first example, but here based on
a linear force-free model. Ambiguities (leading to ghost features) are removed by
using the solution which is closest to the magnetic field model.

\section{Conclusions}
\label{conclusions}
 We describe a newly developed tool for the stereoscopic reconstruction of
 plasma loops from two images. The tool is intended to be used for the
 STEREO-mission. Within this work we tested the method with the help
 of a model active region, from which we computed images from two different
 viewpoints (artificial STEREO images). We tried to reconstruct the original
 3D loops from the artificial images and compared the result with the original.

 As a first step we tried a classical geometric stereoscopic reconstruction.
 The corresponding reconstruction contains ambiguities, because multiple
 points of intersection occur. As a consequence the reconstruction contains
 the correct 3D loops but also additional several so called ghost features.
 Within this approach we cannot distinguish between real and ghost features.

 It is helpful, that the coronal plasma has a high conductivity and
 consequently the emitting plasma also outlines the coronal magnetic
 field. This means that a reconstruction of coronal loops is equivalent
 to the reconstruction of the magnetic field, e.g., a perfect magnetic field
 reconstruction would (if the correct, emitting magnetic field lines are chosen)
 also provide the plasma loops. Unfortunately it is hard to get the accurate
 coronal magnetic field and usually one has to extrapolate it from photospheric
 measurements with e.g., non-linear and linear force-free models or potential fields.
 We investigated how such coronal magnetic field models can be used to associate
 features in the two (artificial) STEREO-images. The method also provides us
 with a quantitative measure of the agreement between the magnetic field model
 and the observed loops.
 To do so we compute the average
 distances of the loops and the projections of magnetic field lines. If
 the distance measure $C$ is sufficiently small then the magnetic field model
 is already a good proxy of the 3D plasma loops. This is the ideal situation. As
 a result we not only get the 3D loops, but as well a reasonable coronal magnetic field
 model.

 Due to noise, ambiguities and limited information (say if we only have
 line-of-sight magnetograms instead of vectormagnetograms) the plasma loops
 and magnetic loops (measured in both 2D projections) do not agree. In this
 case we obtain finite values of the measure $C$ and the magnetic field proxy
 is not a good (or perfect) approximation for the 3D plasma loop. The magnetic
 field proxy can help, however, to eliminate ambiguities in the geometric
 stereoscopic reconstruction. Firstly, the proxy helps to associate features
 in both images with each other and secondly if multiple intersection points
 still occur during the stereoscopy, we choose the point closest to the
 corresponding magnetic field line. Even an imperfect (or even an inconsistent
 one, e.g., different values of $\alpha$ on different loops within the linear
 force-free approach) coronal magnetic field can be used for this aim.

 As an outlook one might think about using the stereoscopic reconstructed
 plasma loops to improve the coronal magnetic field model.

\begin{acknowledgements}
The work of Wiegelmann was supported by DLR-grant 50 OC 0501.
\end{acknowledgements}
\appendix
\section{The model}
\label{appendix1}
We test our stereoscopic reconstruction tools with the help of an model
active region. The advantage is here, that we know the exact solution
and can check if our stereoscopy tools are able to find a reasonable
reconstruction of the original.

We use the semi-analytic non-linear force-free field model developed by
\inlinecite{low:etal90} as a model active region coronal magnetic field, with the
parameters $l=0.5$ and $\Phi=1.4$. We compute this model active region in a $96
\times 96 \times 96$ computational box. To avoid (for the magnetic modelling tools)
boundary effects we display only the center $64 \times 64 $ region in the 2D images
(Figure \ref{fig2}) and the center $64 \times 64 \times 80$ 3D box (Figure
\ref{fig3}).

We use the following spacecraft locations. STEREO-1 is somewhat below the ecliptic
at $(-120,-10,215)$ and STEREO-2 is somewhat above the ecliptic at $(108,10,215)$.
The angle between the spacecraft is $56^o$ To compute artificial STEREO images from
two different viewpoints we fill the magnetic field lines with plasma. To do so we
use the scaling law $F_H \approx B/L$ which has been developed by
\inlinecite{schrijver:etal04} to compute artificial EUV images from a potential
field magnetic field model. We used a somewhat modified approximation $F_H \approx
B/(L+10)$, where the magnetic field is in $G$ and the length in pixel. (An absolute
scaling is not necessary here, because these data are only used to create the
artificial STEREO images.)

 We show the 3D structure
of four coronal loops in Figure \ref{fig1} a). To compute a 3D density distribution
we calculate a bundle of field lines for each loop, the center loop shown in Figure
\ref{fig1} a) and 11 more loops with footpoints located in a circle with the radius
of $0.5$ pixel on the photosphere around the center loops. All loops are filled with
plasma by the scaling law. Figure \ref{fig2} shows two artificial images which mimic
the different views of two STEREO-spacecraft. The artificial images have been taken
with an angle of $56^o$ by a line-of-sight integration. The images show also the
projection of the four center magnetic field lines.

We test our  stereoscopic tools in the sense that we try to reconstruct the 3D
structure of the flux tubes from the loop projection from  two viewpoints shown in
Figure \ref{fig2} and compare the result with the original (center) loops shown in
Figure \ref{fig1} a).
 %


\begin{thebibliography}{32}

\bibitem[\protect\citeauthoryear{Amari, Boulmezaoud and
  Mikic}{1999}]{amari:etal99}
{Amari}, T., {Boulmezaoud}, T.Z. and {Mikic}, Z.: 1999,  {\it Astron. Astrophys.},
{\bf 350}, 1051.

\bibitem[\protect\citeauthoryear{Aschwanden}{2005}]{aschwanden05}
{Aschwanden}, M.J.: 2005,
 {\it Solar Phys.}, {\bf 228}, 339.

\bibitem[\protect\citeauthoryear{Aschwanden et~al.}{1999}]{aschwanden:etal99}
{Aschwanden}, M.J., {Newmark}, J.S., {Delaboudini{\`e}re}, J.-P.,
      {Neupert}, W.M., {Klimchuk}, J.A., {Gary}, G.A.,
      {Portier-Fozzani}, F. and {Zucker}, A.:
  1999,  {\it Astrophys. J.}, {\bf 515}, 842.

\bibitem[\protect\citeauthoryear{Aschwanden et~al.}{2000}]{aschwanden:etal00}
{Aschwanden}, M.J., {Alexander}, D., {Hurlburt}, N.,
      {Newmark}, J.S., {Neupert}, W.M., {Klimchuk}, J.A. and
      {Gary}, G.A.: 2000,  {\it Astrophys. J.}, {\bf 531}, 1129.


\bibitem[\protect\citeauthoryear{Batchelor}{1994}]{batchelor94}
{Batchelor}, D.: 1994,  {\it Solar Phys.}, {\bf 155}, 57.

\bibitem[\protect\citeauthoryear{Berton and Sakurai}{1985}]{berton:etal85}
{Berton}, R. and {Sakurai}, T.: 1985,  {\it Solar Phys.}, {\bf 96}, 93.

\bibitem[\protect\citeauthoryear{Carcedo et~al.}{2003}]{carcedo:etal03}
{Carcedo}, L., {Brown}, D.S., {Hood}, A.W., {Neukirch}, T. and {Wiegelmann}, T.:
2003,  {\it Solar Phys.}, {\bf 218}, 29.

\bibitem[\protect\citeauthoryear{Chiu and Hilton}{1977}]{chiu:etal77}
{Chiu}, Y.T. and {Hilton}, H.H.: 1977,  {\it Astrophys. J.}, {\bf 212}, 873.

\bibitem[\protect\citeauthoryear{Gary, Davis and Moore}{1998}]{gary:etal98}
{Gary}, G.A., {Davis}, J.M. and {Moore}, R.: 1998,  {\it Solar Phys.}, {\bf 183},
45.

\bibitem[\protect\citeauthoryear{Inhester and Wiegelmann}{2006}]{inhester:etal06}
{Inhester}, B. and {Wiegelmann}, T.: 2006, {\it Solar Phys., in press}

\bibitem[\protect\citeauthoryear{Lee, Newman and Gary}{2006}]{lee:etal06}
{Lee}, J.K., {Newman}, T.S. and {Gary}, G.A.: 2006, {\it Pattern Recognition}, {\bf
39}, 246.

\bibitem[\protect\citeauthoryear{Low and Lou}{1990}]{low:etal90}
{Low}, B.C. and {Lou}, Y.Q.: 1990,  {\it Astrophys. J.}, {\bf 352}, 343.

\bibitem[\protect\citeauthoryear{Marsch, Wiegelmann and Xia}{2004}]{marsch:etal04}
{Marsch}, E., {Wiegelmann}, T. and {Xia}, L.D.: 2004,  {\it Astron. Astrophys.},
{\bf 428}, 629.

\bibitem[\protect\citeauthoryear{Portier-Fozzani and Inhester}{2001}]{portier-fozzani:etal01}
{Portier-Fozzani}, F. and {Inhester}, B.: 2001, {\it Space Science Reviews}, {\bf
97}, 51.

\bibitem[\protect\citeauthoryear{R{\' e}gnier, Amari and Kersal{\' e}}{2002}]{regnier:etal02}
{R{\' e}gnier}, S., {Amari}, T. and {Kersal{\' e}}, E.: 2002,  {\it Astron.
Astrophys.}, {\bf 392}, 1119.

\bibitem[\protect\citeauthoryear{Sakurai}{1981}]{sakurai81}
{Sakurai}, T.: 1981,  {\it Solar Phys.}, {\bf 69}, 343.

\bibitem[\protect\citeauthoryear{Schmidt and Bothmer}{1996}]{schmidt:etal96}
{Schmidt}, W.K.H. and {Bothmer}, V.: 1996, {\it Adv. Space Res.}, {\bf 17}, 369.

\bibitem[\protect\citeauthoryear{Schrijver et~al.}{1999}]{schrijver:etal99}
{Schrijver}, C.J., {Title}, A.M., {Berger}, T.E.,
      {Fletcher}, L., {Hurlburt}, N.E., {Nightingale}, R.W.,
      {Shine}, R.A., {Tarbell}, T.D., {Wolfson}, J., {Golub}, L.,
      {Bookbinder}, J.A., {Deluca}, E.E., {McMullen}, R.A.,
      {Warren}, H.P., {Kankelborg}, C.C., {Handy}, B.N. and
      {de Pontieu}, B.:
1999,
 {\it Solar Phys.}, {\bf 187}, 261.

\bibitem[\protect\citeauthoryear{Schrijver et al.}{2004}]{schrijver:etal04}
{Schrijver}, C.J., {Sandman}, A.W., {Aschwanden}, M.J. and {DeRosa}, M.L.:
  2004,  {\it Astrophys. J.}, {\bf 615}, 512.


\bibitem[\protect\citeauthoryear{Seehafer}{1978}]{seehafer78}
{Seehafer}, N.: 1978,  {\it Solar Phys.}, {\bf 58}, 215.

\bibitem[\protect\citeauthoryear{Semel}{1967}]{semel67}
{Semel}, M.: 1967, {\it Annales d'Astrophysique}, {\bf 30}, 513.

\bibitem[\protect\citeauthoryear{Valori, Kliem and Keppens}{2005}]{valori:etal05}
{Valori}, G., {Kliem}, B. and {Keppens}, R.: 2005,  {\it Astron. Astrophys.}, {\bf
433}, 335.

\bibitem[\protect\citeauthoryear{Wheatland}{2004}]{wheatland04}
{Wheatland}, M.S.: 2004,  {\it Solar Phys.}, {\bf 222}, 247.

\bibitem[\protect\citeauthoryear{Wheatland, Sturrock and Roumeliotis}{2000}]{wheatland:etal00}
{Wheatland}, M.S., {Sturrock}, P.A. and {Roumeliotis}, G.: 2000,  {\it Astrophys.
J.}, {\bf 540},
  1150.

\bibitem[\protect\citeauthoryear{Wiegelmann}{2004}]{wiegelmann04}
{Wiegelmann}, T.: 2004,  {\it Solar Phys.}, {\bf 219}, 87.

\bibitem[\protect\citeauthoryear{Wiegelmann and Inhester}{2003}]{wiegelmann:etal03a}
{Wiegelmann}, T. and {Inhester}, B.: 2003,  {\it Solar Phys.}, {\bf 214}, 287.

\bibitem[\protect\citeauthoryear{Wiegelmann and Neukirch}{2002}]{wiegelmann:etal02}
{Wiegelmann}, T. and {Neukirch}, T.: 2002,  {\it Solar Phys.}, {\bf 208}, 233.

\bibitem[\protect\citeauthoryear{Wiegelmann and Neukirch}{2003}]{wiegelmann:etal03}
{Wiegelmann}, T. and {Neukirch}, T.: 2003, {\it Nonlinear Processes in Geophysics},
{\bf 10},
  313.

\bibitem[\protect\citeauthoryear{Wiegelmann, Inhester and Sakurai}{2006}]{wiegelmann:etal06}
{Wiegelmann}, T., {Inhester}, B. and {Sakurai}, T.: 2006,  {\it Solar Phys.}, {\bf
233}, 215.

\bibitem[\protect\citeauthoryear{Wiegelmann et~al.}{2005a}]{wiegelmann:etal05b}
{Wiegelmann}, T., {Inhester}, B., {Lagg}, A. and {Solanki}, S.K.:
  2005a, {\it Solar Phys.}, {\bf 228}, 67.

\bibitem[\protect\citeauthoryear{Wiegelmann et~al.}{2005b}]{wiegelmann:etal05}
{Wiegelmann}, T., {Lagg}, A., {Solanki}, S.K., {Inhester}, B. and {Woch}, J.:
  2005b,  {\it Astron. Astrophys.}, {\bf 433}, 701.

\bibitem[\protect\citeauthoryear{Yan and Sakurai}{2000}]{yan:etal00}
{Yan}, Y. and {Sakurai}, T.: 2000,  {\it Solar Phys.}, {\bf 195}, 89.

\end{thebibliography}

\end{article}
\end{document}